\documentclass{article}

\usepackage{arxiv}

\usepackage[utf8]{inputenc} % allow utf-8 input
\usepackage[T1]{fontenc}    % use 8-bit T1 fonts
\usepackage{hyperref}       % hyperlinks
\usepackage{url}            % simple URL typesetting
\usepackage{booktabs}       % professional-quality tables
\usepackage{amsfonts}       % blackboard math symbols
\usepackage{nicefrac}       % compact symbols for 1/2, etc.
\usepackage{microtype}      % microtypography
\usepackage{lipsum}		% Can be removed after putting your text content
\usepackage{graphicx}
\usepackage{natbib}
\usepackage{doi}

\usepackage{amsmath}
\usepackage{amssymb}
\usepackage{algorithm}
\usepackage{algpseudocode}
\usepackage{tikz}
\usepackage{graphicx,subfigure}
\usepackage{makecell}
\usetikzlibrary{shapes.arrows}
\usetikzlibrary{calc}
\usepackage{array,multirow}
\DeclareMathOperator*{\minimize}{minimize}
\DeclareMathOperator*{\argmin}{argmin}
\usepackage{multicol}

\title{Accurate and Interpretable Solution of the Inverse Rig for Realistic Blendshape Models with Quadratic Corrective Terms}

\date{November 2022}	% Here you can change the date presented in the paper title
\author{ \href{https://orcid.org/0000-0002-5656-9189}{\includegraphics[scale=0.06]{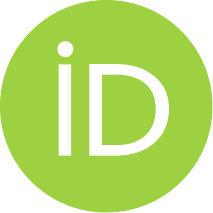}\hspace{1mm}Stevo Racković} \\
	Department of Mathematics\\
	Instituto Superior Técnico\\
	Lisbon, Portugal \\
	\texttt{stevo.rackovic@tecnico.ulisboa.pt} \\
	\And
	\href{https://orcid.org/0000-0003-3071-6627}{\includegraphics[scale=0.06]{orcid.pdf}\hspace{1mm}Cláudia Soares} \\
	Department of Computer Sceince\\
	NOVA School of Science and Technology\\
	Caparica, Portugal \\
	\And
	\href{https://orcid.org/0000-0003-3497-5589}{\includegraphics[scale=0.06]{orcid.pdf}\hspace{1mm}Dušan Jakovetić} \\
	Department of Mathematics\\
	University of Novi Sad\\
	Novi Sad, Serbia \\
 	\And
	Zoranka Desnica \\
	3Lateral Animation Studio\\
	Epic Games Company \\
}

\renewcommand{\shorttitle}{ }

\hypersetup{
pdftitle={Accurate and Interpretable Solution of the Inverse Rig for Realistic Blendshape Models with Quadratic Corrective Terms},
pdfauthor={Stevo Racković, Cláudia Soares, Dušan Jakovetić, Zoranka Desnica},
pdfkeywords={Inverse Rig, Quadratic blendshape model, Majorization Minimization},
}

\begin{document}
\maketitle

\begin{figure*}
  \begin{tikzpicture}
    \node[above right, inner sep=0] (image) at (0,0){\includegraphics[width=\textwidth]{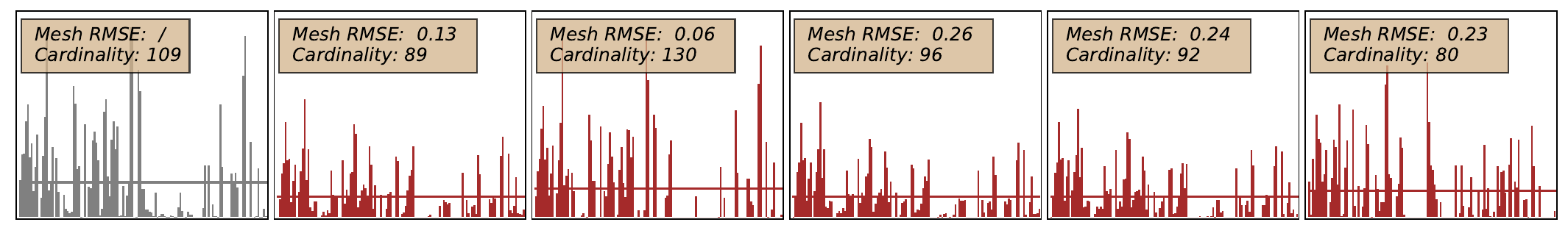}};
    \node[above right, inner sep=0] (image2) at (0,2.7){\includegraphics[width=\textwidth]{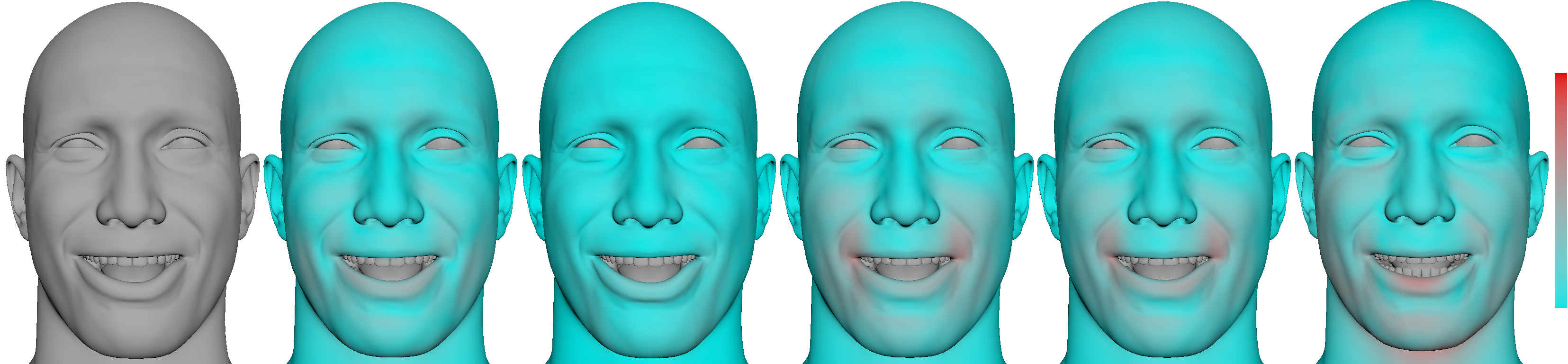}};
    \begin{scope}[
        x={($0.1*(image.south east)$)},
        y={($0.1*(image.north west)$)}]
        \node[darkgray] at  (0.90,0.){\small Reference };
        \node[darkgray] at  (2.60,0.){\small MM (ours) };
        \node[darkgray] at  (4.20,0.){\small SQP (ours)};
        \node[darkgray] at  (5.85,0.){\small Cet };
        \node[darkgray] at  (7.50,0.){\small Cet - loc };
        \node[darkgray] at  (9.10,0.){\small Seol };
            \node[darkgray] at  (9.9,24.3){\footnotesize .49 };
            \node[darkgray] at  (9.9,12.0){\footnotesize .00 };
            \node[darkgray] at  (9.9,11.0){\footnotesize cm };
    \end{scope}
  \end{tikzpicture}
  \caption{The reference frame mesh and the estimates of the approach that solves the proposed objective function using our novel algorithm (\textit{MM}), the approach that solves the proposed objective applying a general-purpose solver (\textit{SQP}), and linear approaches proposed by \cite{cetinaslan2020sketching} (\textit{Cet} for the standard  case, and \textit{Cet-loc} for localized approximation) and \cite{seol2011artist} (Seol). The top row shows obtained meshes, while the bottom represents corresponding activations of the blendshape weights. Red tones in the meshes indicate a higher error of the fit, according to the color bar on the right. The average weight activation of each solution is indicated with a horizontal line. The average mesh error and cardinality (i.e., the number of non-zero weights) of the weight vector are given in a text box for each method --- we aim for the lowest error while keeping the cardinality relatively low. \textit{SQP} yields the cleanest mesh fit of all, yet the cardinality of the weight vector is too high, hence it might compromise the stability of the solution. On the other side, linear methods \textit{Cet, Cet-loc} and \textit{Seol} give visible flaws in mesh reconstruction. \textit{MM} is the only method that provides both an accurate mesh fit and a stable solution.}
  \label{fig:teaser}
\end{figure*}

\begin{abstract}
    We propose a new model-based algorithm for solving the inverse rig problem in facial animation retargeting, exhibiting higher accuracy of the fit and sparser, more interpretable weight vector compared to state-of-the-art methods. The proposed method targets a specific subdomain of human face animation --- highly-realistic blendshape models used in the production of movies and video games. In this paper, we formulate an optimization problem that takes into account all the requirements of targeted models. Unlike the prior solutions, our objective goes beyond a linear blendshape model and employs the quadratic corrective terms necessary for correctly fitting fine details of the mesh. Further, we show that the solution to the proposed problem yields highly accurate mesh reconstruction even when general-purpose solvers, like sequential quadratic programming, are used. The results obtained using general-purpose solvers are highly accurate in the mesh space but do not exhibit favorable qualities in terms of weight sparsity and smoothness, and for this reason, we further propose a novel algorithm relying on a majorization-minimization technique. The algorithm is specifically suited for solving the proposed objective, yielding a high-accuracy mesh fit while respecting the constraints and producing a sparse and smooth set of weights that are easy to manipulate and interpret by animation artists. We show results using both proprietary and open-source animated characters of high quality and level of detail. Our algorithm is benchmarked with state-of-the-art approaches, and shows an overall superiority of the results, yielding a smooth animation reconstruction with a relative improvement up to $45\%$ in root mean squared mesh error while keeping the cardinality comparable with benchmark methods. This paper gives a comprehensive set of evaluation metrics that cover different aspects of the solution, including mesh accuracy, sparsity of the weights, and smoothness of the animation curves, as well as the appearance of the produced animation, which human experts evaluated.
\end{abstract}

% keywords can be removed

% ####################################################################################################################

\section{Introduction}

Facial animation is a growing research topic in academia as well as in industry due to the central role of facial expressions in verbal and non-verbal communication, impacting mainly the arts (animation, video games) and other areas, like marketing (animated advertisement, video chatbots). Although there is a number of models developed to credibly deform a 3D character's face, including the underlying anatomical structure \cite{choe2001performance, sifakis2005automatic, wu2016anatomically, zoss2018empirical, zoss2019accurate, Fratarcangeli2020FastNL}, physics-based deformations \cite{hahn2013efficient, ichim2016building, ichim2017phace}, or morphable models \cite{hu20053d, smet2010optimal, thies2015realtime, wang2020facial}, the most popular approach is the blendshape model \cite{Pighin1998SynthesizingRF, choe2001analysis, joshi2006learning, deng2006animating, lewis2010direct, li2010example, lewis2014practice}. Even though this approach might suffer from reduced expressivity, it provides intuitive controls and is easy to use and understand \cite{ichim2017phace}. Traditionally, the (delta) blendshape model is presented as a linear mapping 
\begin{equation}\label{eq:linear_bs_funciton}
    f_L(\textbf{w}) = \textbf{b}_0 + \sum_{i=1}^m w_i\textbf{b}_i = \textbf{b}_0 + \textbf{Bw},
\end{equation}
where $\textbf{b}_0\in\mathbb{R}^{3n}$ is a column vector representing the face mesh (with $n$ vertices) in a resting position and  $\textbf{b}_1,...,\textbf{b}_m\in\mathbb{R}^{3n}$ are delta blendshape vectors, i.e., these are topologically identical copies of $\textbf{b}_0$, but each of them is locally deformed, corresponding to an atomic facial expression. Blendshape vectors span the space of feasible expressions of the 3D character and can be invoked isolated or combined with other blendshapes to build more complex facial expressions. This is achieved by assigning corresponding activation weights $w_1,...,w_m$ to each blendshape. These weights are scalars, and are often (but not exclusively) restricted to the $[0,1]$ interval. To ease the notation, blendshape weights can be concatenated into a column weight vector $\textbf{w}=[w_1,...,w_m]^T$, and blendshape vectors can be collected to build columns of a blendshape matrix $\textbf{B}=[\textbf{b}_1,...,\textbf{b}_m]$, which gives a matrix expression in (\ref{eq:linear_bs_funciton}). (More details will be given in Sec. \ref{ss:linear_model})

\begin{figure}
    \centering
    \includegraphics[width=0.4\linewidth]{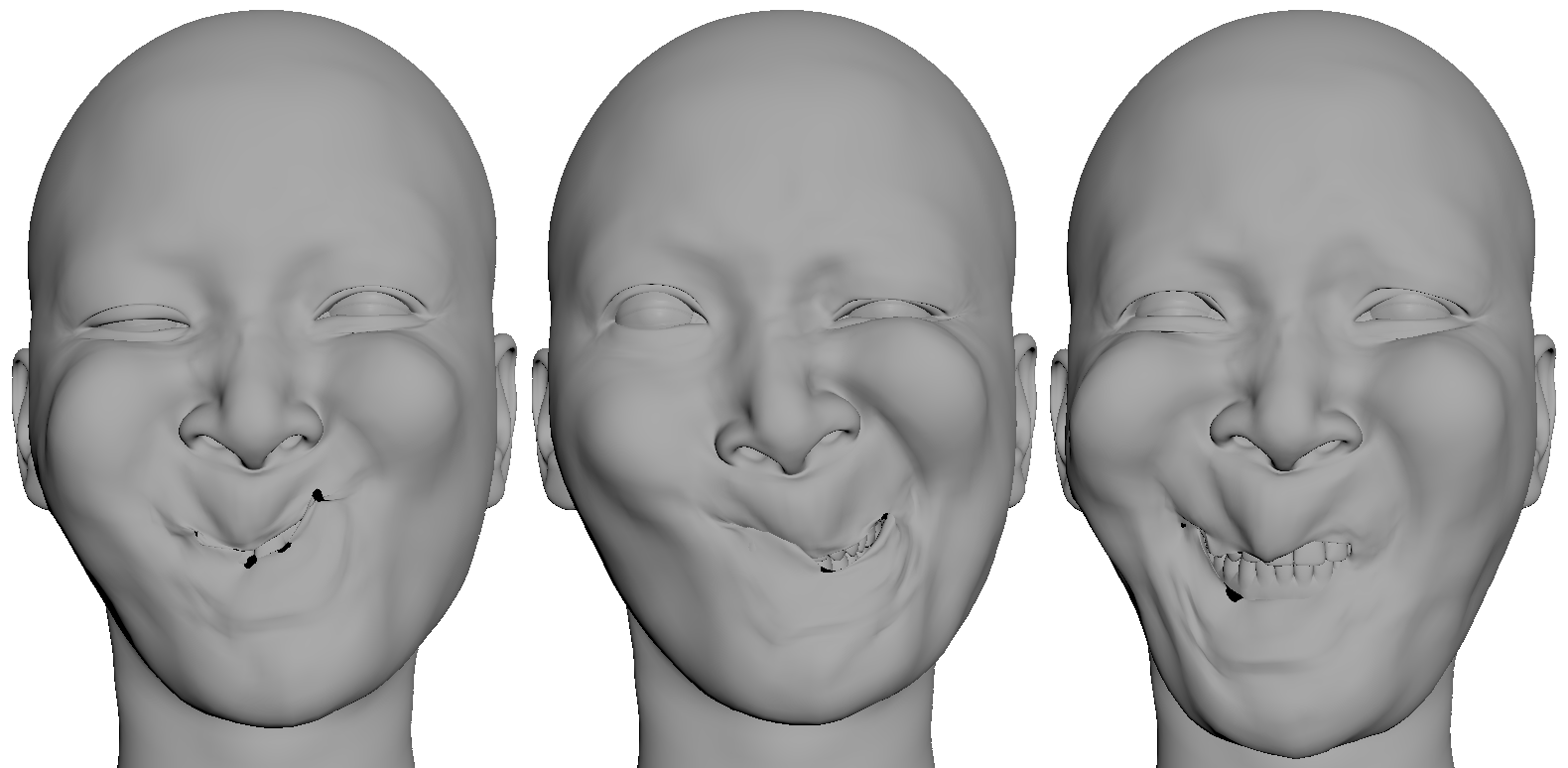}
    \caption{Examples of dense activation vectors. All the activation weights of this character model are set to a random value within a feasible interval $[0,1]$, and we can see that there are many anatomically incorrect deformations as well as the breaking of the mesh.}
    \label{fig:artifacts}
\end{figure}

Once the blendshape model is built, a character can be animated directly by setting the values of the weights for each animation frame. Since high-quality characters usually have over one hundred blendshapes, manually setting the weights for each frame involves intensive labor, and large production times. The animation can be automated if the reference motion is available in the form of 3D scans or a set of markers obtained via motion capture (MoCap). This process is called animation retargeting, and algorithms are developed to estimate the optimal set of activation weights for each frame so that the character mesh is deformed to closely resemble a reference mesh $\widehat{\textbf{b}}\in\mathbb{R}^{3n}$ at a given frame. We call the problem of finding an optimal $\textbf{w}$ the \textbf{inverse rig problem}. This solution vector needs to satisfy several properties in order to be considered good. In the first place, we measure the data fidelity, i.e., the resemblance between a reconstructed and a reference mesh. Another essential feature is the stability of the solution, so that if the few activated weights are slightly adjusted afterwards, the mesh behaves predictably and does not produce artifacts like discontinuities or anatomically impossible positions (Figure \ref{fig:artifacts}). This is important because the results of the automatic retargeting are often adjusted manually in production \cite{cetinaslan2020stabilized}. For the same reason, it is preferable to have lower cardinality, i.e., the number of non-zero values of the weight vector. If the animation retargeting is performed over a continuous time sequence, there should not be visible discontinuities between the frames. Finally, if the range of weight values is restricted by the animation framework in use (e.g., $0\leq w_i\leq 1$), the solution needs to respect the constraints.

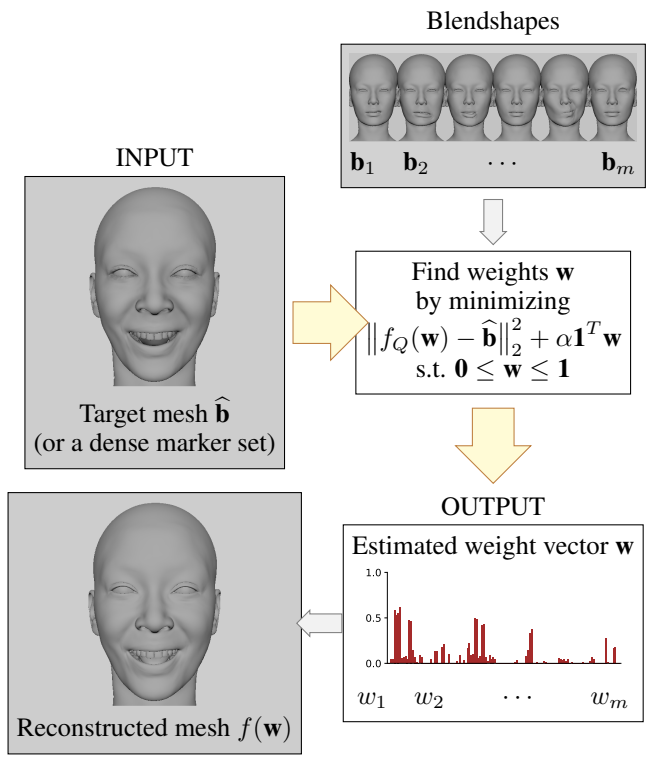
\begin{figure}
\centering
    \begin{tikzpicture}
        \tikzset{vertex/.style = {shape=rectangle,draw,minimum size=1.5em}}
        \tikzset{edge/.style = {->,> = latex'}}
        \node[vertex,label=INPUT,fill=gray!39] (F1) at  (0,4) {\makecell{\includegraphics[width=.7in]{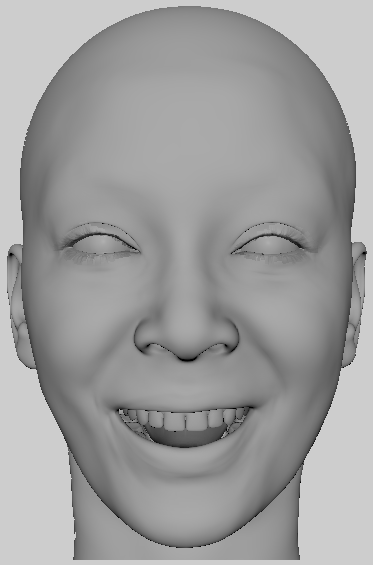}\\ Target mesh $\widehat{\textbf{b}}$ \\ (or a dense marker set)}};
        \node[vertex,label=Blendshapes,fill=gray!35] (F2) at  (4.5,6.75) {\makecell{\includegraphics[width=1.5in]{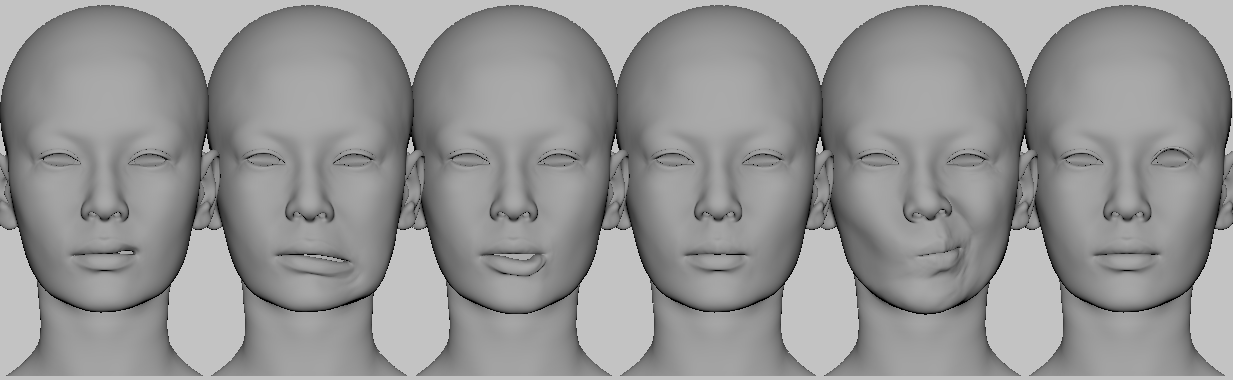} 
        \\ $\textbf{b}_1 \quad \textbf{b}_2 \qquad \cdots \qquad\quad \textbf{b}_m$}};
        \node[vertex] (Eq) at  (4.5,4) {\makecell{ Find weights $\textbf{w}$ \\ by minimizing \\ 
        $\big\|f_Q(\textbf{w})-\widehat{\textbf{b}}\big\|_2^2 + \alpha \textbf{1}^T\textbf{w}$ \\ s.t. $\textbf{0} \leq \textbf{w} \leq \textbf{1}$}};
        \node[vertex,label=OUTPUT] (F3) at  (4.5,0) {\makecell{ Estimated weight vector $\textbf{w}$ \\ \includegraphics[width=1.4in]{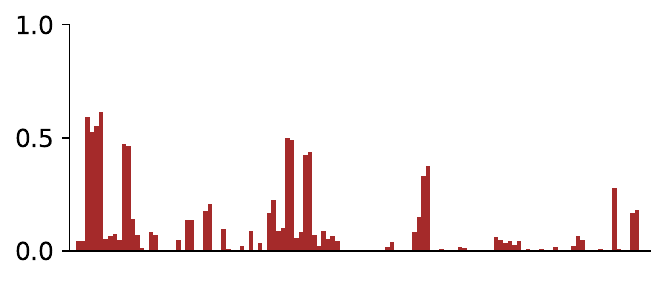} \\ $w_1 \quad w_2 \qquad \cdots \qquad w_m$}};
        \node[vertex,fill=gray!39] (F4) at  (0,0) {\makecell{\includegraphics[width=.7in]{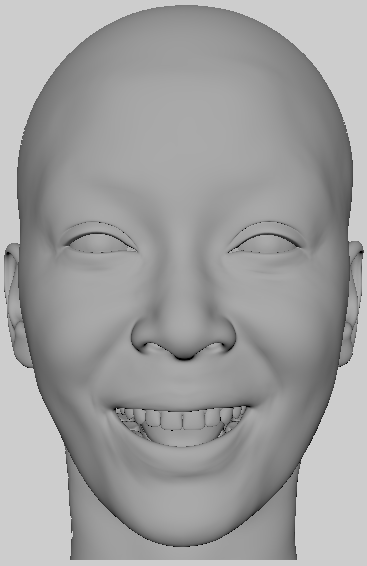}\\ Reconstructed mesh $f(\textbf{w})$}};
        
        \node[single arrow, draw=brown, fill=yellow!20, 
              minimum width = 10mm, single arrow head extend=1mm,
              minimum height=10mm] at (2.2,4) {} ;
        \node[single arrow, draw=gray, fill=gray!10, 
              minimum width = 2mm, single arrow head extend=0.5mm,
              minimum height=6mm, rotate=-90] at (4.5,5.40) {} ;
        \node[single arrow, draw=brown, fill=yellow!20, 
              minimum width = 10mm, single arrow head extend=1mm,
              minimum height=10mm, rotate=-90] at (4.5,2.5) {} ;
        \node[single arrow, draw=gray, fill=gray!10, 
              minimum width = 2mm, single arrow head extend=0.5mm,
              minimum height=6mm, rotate=180] at (2.25,0) {} ;
    \end{tikzpicture}
\caption{A schematic representation of the considered problem. Some reference mesh $\widehat{\textbf{b}}$ is taken as an input, and the algorithm needs to estimate the activation weights of the quadratic blendshape model of the character in order to closely reconstruct the reference mesh and respect the structure of the model. The optimization problem might be solved using general-purpose quadratic solvers, like \cite{byrd1999interior}, or using the algorithm proposed in this paper. Further, the estimated weights are plugged-in into the animation software to give the final reconstructed expression.}\label{fig:scheme}
\end{figure}

Our main interest in this paper is a solution to the inverse rig problem for realistic human characters (Figure \ref{fig:scheme}). We make a distinction between \textit{data-based} and \textit{model-based} approaches for solving the inverse rig --- the former assumes a rig function as a black box and demands long animation sequences that span a whole space of expected expressions in order to train a good regressor, while the latter only requires a well-defined rig function with corresponding basis vectors. The literature offers several model-based methods for solving the inverse rig under the linear blendshape model \cite{choe2001analysis, joshi2006learning, lewis2010direct, seol2011artist, Liu2010LocalizedOF, cetinaslan2020sketching}; however, due to the increasing complexity and level of realism of the avatars in the movie and gaming industry (but also for purposes of communication, education, virtual reality), linear models do not provide a high-enough level of detail. A possible approach is the application of data-based machine learning algorithms, yet this is an expensive alternative as it demands a large amount of data to provide a good fit. 

In this paper, we propose a model-based algorithm that solves the inverse rig problem for realistic human face blendshape models used in the industry, taking into account the quadratic corrective terms of a blendshape model and the constraints over the weights vector. Our algorithm is benchmarked with state-of-the-art methods \cite{seol2011artist, cetinaslan2020sketching}, and it exhibits a significant advantage in data fidelity of the reconstructed meshes (a relative improvement up to $45\%$ in root mean squared mesh error, computed as explained in Section \ref{ss:metrics}) while at the same time cardinality (i.e., the number of non-zero elements) of the weight vector is comparable to previous solutions, and the frame-to-frame transitions are smooth. 

% -------------------------------------------------------------------------------------------------------------------
\subsection{Contributions}

The contributions of this paper are two-fold. 
\begin{itemize}
    \item We address the inverse rig problem with quadratic corrective terms \cite{seo2011compression, lewis2014practice}, and show evidence of decreasing mesh error when compared with the commonly used linear blendshape model (\ref{eq:linear_bs_funciton}). To the best of our knowledge, this paper presents the first model-based solution to the inverse rig problem under the quadratic blendshape function. We show experimentally that the new objective can be optimized using standard solvers with significantly reduced error in the mesh space compared to the state-of-the-art methods. Specifically, we use an interior-point-based solver within Python library \texttt{scipy} \cite{2020SciPy}, that is based on the sequential Quadratic Programming (SQP) applied iteratively with Trust-Region method \cite{byrd1999interior, conn2000trust}. However, using such a general-purpose solver empirically shows an increased number of activated blendshapes and frame-to-frame discontinuities. This leads us to our second contribution.
    \item This paper formulates an algorithm that is specifically suited for solving the inverse rig problem for a quadratic blendshape model. (Details of the quadratic blendshape model are given in Section \ref{ss:quadratic_model}.) In particular, we apply the majorization-minimization technique \cite{sun2016majorization,marnissi2020majorize} and devise a surrogate function that allows for an efficient iterative solution to the proposed optimization problem. The obtained solution has a low mesh error, while simultaneously decreasing the number of activated blendshape weights (cardinality). Frame-to-frame transitions are shown to be smooth, and a user study with animation artists has evidenced the higher quality of our approach.
\end{itemize}

This paper presents the solution from a domain point of view, including an explanation of how the method works, a comprehensive set of experiments on proprietary and open-source state-of-the-art animation characters, a detailed modeling rationale and motivation for introducing and modeling quadratic blendshape models, the intuition behind the method and implementation details. In a companion paper \cite{rackovic2022majorization}, we present the detailed method derivation from the optimization theory perspective and convergence analysis of the algorithm.

The rest of this paper is organized as follows. In Section 2, we cover the prior work and relate our method to the corresponding directions of research. In Section 3, we cover the main concepts of the blendshape model for facial animation and the problem of rig inversion. Section 4 introduces the proposed method for solving the inverse rig, and Section 5 gives an extensive numerical evaluation of the proposed method. Finally, Section 6 concludes the paper. 

% -------------------------------------------------------------------------------------------------------------------
\subsection{Notation}

Throughout this paper, scalar values will be denoted with lowercase Latin $a,b,c$, or lowercase Greek $\alpha,\beta,\gamma$ letters. Vectors are represented with bold lowercase $\textbf{a,b,c}$ and are indexed using a subscript, i.e., the $i^{th}$ element of the vector $\textbf{a}$ is $a_i$. If there is a subscript and the letter is still in bold, it is not indexing --- we will use this to differentiate blendshape vectors ($\textbf{b}_0,\textbf{b}_1,...,\textbf{b}_m$) as they have similar properties, or to indicate that a vector $\textbf{a}$ takes specific value at iteration $i$ of an iterative algorithm, which is denoted by a subscript within the brackets $\textbf{a}_{(i)}$. We use $\textbf{0}$ and $\textbf{1}$ to denote vectors of all zeros and all ones, respectively. When we use order relations ($\geq,\leq,=$) between two vectors, it is assumed component-wise. All the vectors are assumed to be column vectors, and $[a_1,...,a_n]^T$ represents a column vector obtained by stacking $n$ scalars.
Matrices are written in bold capital letters $\textbf{A, B, C}$ and also indexed using subscripts --- $\textbf{A}_i$ is the $i^{th}$ row of a matrix $\textbf{A}$, and $A_{ij}$ is an element of a matrix $\textbf{A}$ in a row $i$ and a column $j$. If a superscript is given within the brackets $\textbf{A}^{(i)}$ it denotes a specific matrix corresponding to the (vertex) position $i$. A notation $\textbf{A}=[\textbf{a}_1,...,\textbf{a}_n]$ means that a matrix $\textbf{A}$ is obtained by stacking vectors $\textbf{a}_1,...,\textbf{a}_n$ along the columns.
Functions are given using lowercase Latin or Greek characters, but always with corresponding parameters in the brackets $f(\cdot),g(\cdot),\phi(\cdot),\psi(\cdot)$. A set of real numbers and a set of positive integers are given by $\mathbb{R}$ and $\mathbb{N}$, respectively. The Euclidean norm is denoted by $\|\cdot\|,$ and the L1 norm by $\|\cdot\|_1.$

% ####################################################################################################################

\section{Related Work}

Blendshape animation has been a research topic for more than two decades \cite{Pighin1998SynthesizingRF, choe2001analysis, choe2001performance}. The main tasks in terms of intensive manual labor and extension of production time are (1) the creation of the blendshapes, and (2) the actual animation of the blendshape basis (our work). The main components of a blendshape model are a neutral face mesh and a blendshape basis (local deformations of a neutral face), and there is a body of work proposing automated solutions for creating blendshapes. Two main approaches are (1) building a basis from a dense set of captured data \cite{smet2010optimal, neumann2013sparse, chaudhuri2020personalized, wang2020facial} or (2) deforming a generic set of blendshapes to produce personalized blendshape meshes \cite{li2010example, li2013realtime, ribera2017facial, zhang2020facial, han2021generate, seol2016creating}. In this paper, we assume that the blendshapes are already available and that they closely resemble the actor or user.

The main focus of our paper is to solve the inverse rig problem to produce animation, i.e., to automatically adjust the activation weights so that the resulting mesh follows a reference motion. Reference motion is a (sparse or dense) set of markers recorded from an actor's face using motion capture (MoCap) systems \cite{deng2006animating, seol2014tuning}. A sparse set of markers is a common approach, particularly if the motion should be retargeted to a fantasy character with a face significantly different from the source actor \cite{sifakis2005automatic, hirose2012creating, seol2012spacetime, ouzounis2017kernel, serra2018easy}, and it demands special care in positioning the markers on both source and target faces \cite{reverdy2015optimal}. Although this technique is sufficient for general-purpose MoCap, it fails to capture fine details of the face \cite{furukawa2009dense, reverdy2015optimal}. For this reason, markerless methods are developed to provide high-fidelity performance capture \cite{bradley2010hihg, beeler2011highquality, thies2015realtime}.

The approaches to solving the inverse rig problem can be divided into data-based (regression models that demand long animated sequences for the training phase) and model-based (that do not demand animation for training, only the rig function with the basis vectors). Data-based solutions are popular due to their ability to provide accurate solutions even for complex rig functions, and commonly apply neural networks \cite{holden2016learning, bailey2020fast, song2020accurate, Kim2021DeepLU}, radial basis functions  \cite{deng2006animating, Song2011CharacteristicFR, seol2014tuning, holden2015learning} or other forms of regression \cite{feng2008realtime, yu2014regression, buoaziz2013online}. However, the data acquisition may be too expensive, which is why we consider a model-based approach in this paper. Within model-based approaches, the literature examines only simplified linear blendshape models to fit the acquired mesh, yielding convex optimization problems easy to solve with a closed form \cite{choe2001analysis, sifakis2005automatic, ccetinaslan2016position, li2010example, seol2011artist}. In contrast, we specifically target realistic facial animation with a high level of detail, and for this reason, we need to go beyond the linear model and include quadratic corrective terms, as studied in \cite{holden2016learning, song2017sparse, kim2021optimizing}. 

In this paper, we are not concerned with real-time execution but aim for a more precise mesh fit, hence we assume models with a large number of vertices (we will consider 4000 vertices for each animated character in our experiments).
A different line of work addressing the problem of rig inversion is a direct manipulation of the mesh. It aims for the algorithms and tools that are appended to a sculpted character, allowing an artist to refine the pose by dragging specific vertices of the face directly and producing the desired expression \cite{zhang2004spacetime, lewis2010direct, seo2011compression, cetinaslan2020sketching, cetinaslan2020stabilized}. In this case, it is important that the animator receives feedback during the dragging, hence the optimization takes into account only a sparse set of markers or even a single vertex \cite{lewis2010direct, anjyo2012practical}.  

Despite our formulation being parallelizable, we do not focus on distributed models here, and all the experiments in this paper are performed sequentially over a non-segmented face mesh. Nevertheless, our method is parallelizable, yielding a problem that is separable by components, hence the computations might be distributed to the level of blendshapes. Another approach toward distributed inverse rig solvers is via using face segmentation or clustering. It allows different face regions to be observed and processed independently or in parallel. Early works consider a simple split of the face into upper and lower sets of markers \cite{choe2001analysis}. More recent papers model  complex splits, either manually \cite{seol2011artist, Liu2010LocalizedOF}, semi-automatically \cite{na2011local, tena2011interactive, Fratarcangeli2020FastNL} or automatically \cite{james2005skinning, joshi2006learning, hirose2012creating, reverdy2015optimal, song2017sparse, bailey2020fast}. Clustering based on the underlying deformation model has been considered in \cite{romeo2020data} and \cite{rackovic}, where the goal of the former was to add a secondary motion to an animated character, and the latter proposes a segmentation for solving the inverse rig locally in a distributed fashion.

All the methods available in the literature solve the inverse rig problem using a linear blendshape function. It turns out it is easy and fast to work with. However, it is of significant interest to work with more complex face models that closely resemble a source actor --- a linear model does not exhibit high-enough accuracy for this purpose. In this paper, we introduce a method that allows the application of a more complex blendshape function, that includes quadratic corrective terms, to produce an accurate and sparse solution to the inverse rig problem.

Further details on the principles of blendshape animation can be found in references \cite{lewis2014practice} and \cite{ccetinaslan2016position}.
 
% ####################################################################################################################

\section{Background on Rig Approximation and Inverse Rig}\label{sec:rig_approximation}

In this section, we give a concise presentation on the main principles of blendshape animation. Section \ref{ss:linear_model} introduces the linear delta blendshape model. Section \ref{ss:quadratic_model} introduces quadratic corrective terms that are added on top of a linear model in order to increase the mesh fidelity of realistic human characters. Finally, Section \ref{ss:inverse_rig} explains how inverse rig problems have been formulated and solved according to existing literature.

% -------------------------------------------------------------------------------------------------------------------
\subsection{Linear Blendshape Model}\label{ss:linear_model}

\begin{figure}
    \centering
    \includegraphics[width=0.4\linewidth]{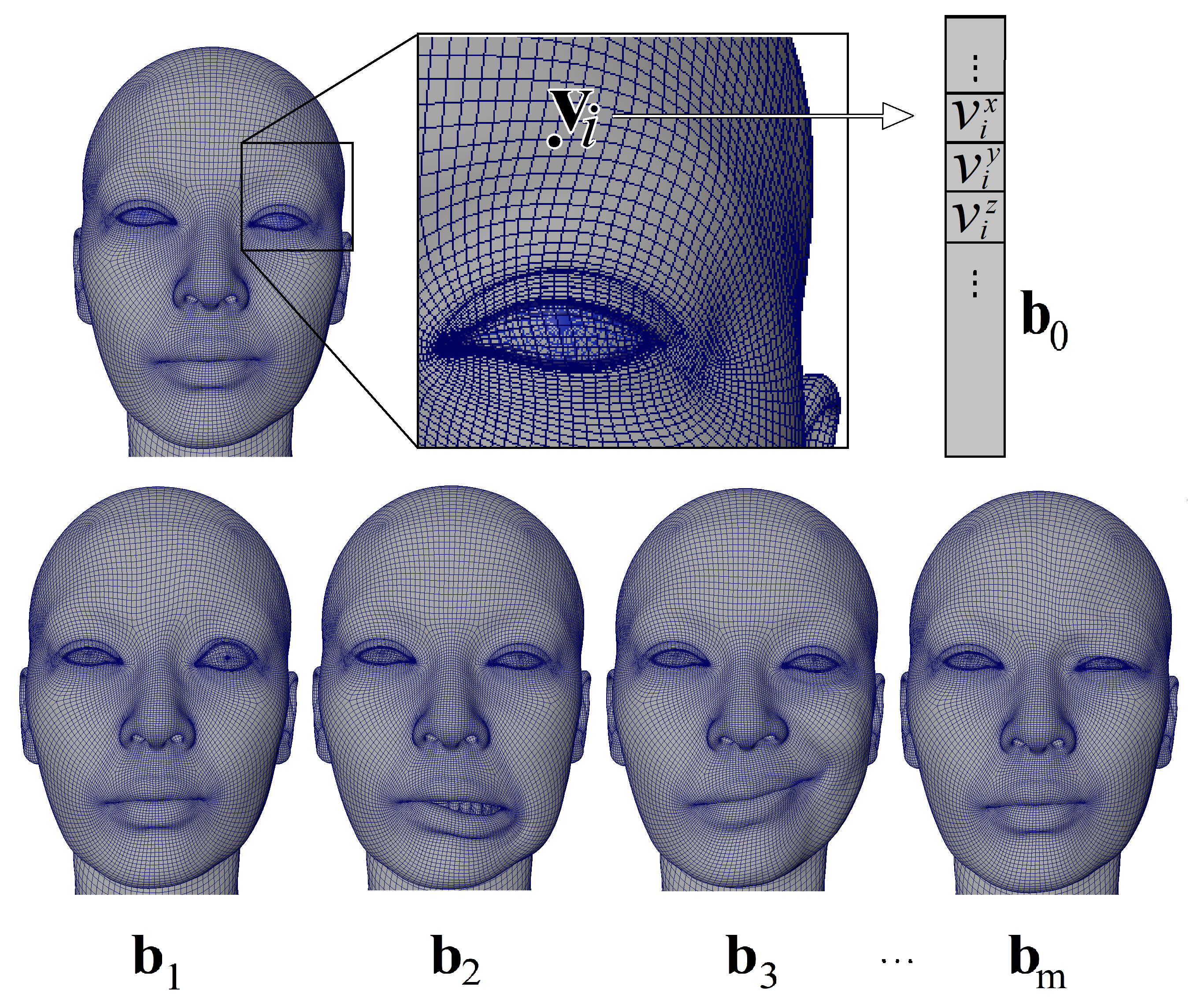}
    \caption{Vectorization of meshes. Neutral mesh $\textbf{b}_0$ on top, and example blendshapes below. Each face vertex $\textbf{v}_i$ for $i=1,...,n$ is unraveled into a vector of coordinates $x,y,z$, and those coordinate vectors are stacked into a single blendshape vector.}
    \label{fig:mesh_vectorization}
\end{figure}

Traditionally, a \textit{blendshape model} consists of a neutral face, that is represented by a column vector $\textbf{b}_0\in\mathbb{R}^{3n}$, and a set of $m$ blendshape vectors $\textbf{b}_1,...,\textbf{b}_m\in\mathbb{R}^{3n}$ that correspond to atomic expressions obtained by local deformations over $\textbf{b}_0$ (Figure \ref{fig:mesh_vectorization}). Each blendshape $\textbf{b}_i$ is assigned an activation parameter $w_i$ that usually (but not exclusively) takes values between $0$ and $1$ \cite{seol2012spacetime}. It is common to use a delta formulation of the blendshape model, where the elements of the blendshape vectors $\textbf{b}_1, ..., \textbf{b}_m$ are not the actual coordinates of the deformed face, but the offsets from their corresponding positions in the neutral $\textbf{b}_0$. (For this reason, some authors use the notation $\Delta\textbf{b}_i$ instead of $\textbf{b}_i$; however, we will not use the $\Delta$ symbol in order to ease the notation.) A \textit{linear delta blendshape function} $f_L(\cdot): \mathbb{R}^m\rightarrow\mathbb{R}^{3n}$ maps activation weights $w_1,...,w_m$ onto the mesh space, and it is defined as 
\begin{equation}
    f_L(w_1,...,w_m) = \textbf{b}_0 + \sum_{i=1}^m w_i\textbf{b}_i.
\end{equation}
If we collect blendshape vectors into a matrix $\textbf{B}=[\textbf{b}_1,...,\textbf{b}_m]$, $\textbf{B}\in\mathbb{R}^{3n\times m}$, the function can be written in a matrix form as 
\begin{equation}\label{eq:linear_matrix_function}
    f_L(\textbf{w}) = \textbf{b}_0 +\textbf{Bw},
\end{equation}
where $\textbf{w} = [w_1,...,w_m]^T$ represents a column vector of blendshape weights.

% -------------------------------------------------------------------------------------------------------------------
\subsection{Quadratic Blendshape Model}\label{ss:quadratic_model}

In modern animation, with an increasing level of detail and with avatars that closely resemble an actor (or a user), linear models are too restrictive and fail to span a desired space of motion. For this reason, additional \textit{corrective shapes} (also known as combination shapes) are included \cite{seo2011compression, lewis2014practice}, and these are usually more numerous than the base vectors. In particular, the quadratic corrective terms are very common, and adding them on top of a linear function (\ref{eq:linear_matrix_function}) significantly improves the accuracy of the representation; hence, we introduce a quadratic blendshape model in the following lines. 

\begin{figure}
    \centering
    \includegraphics[width=0.5\linewidth]{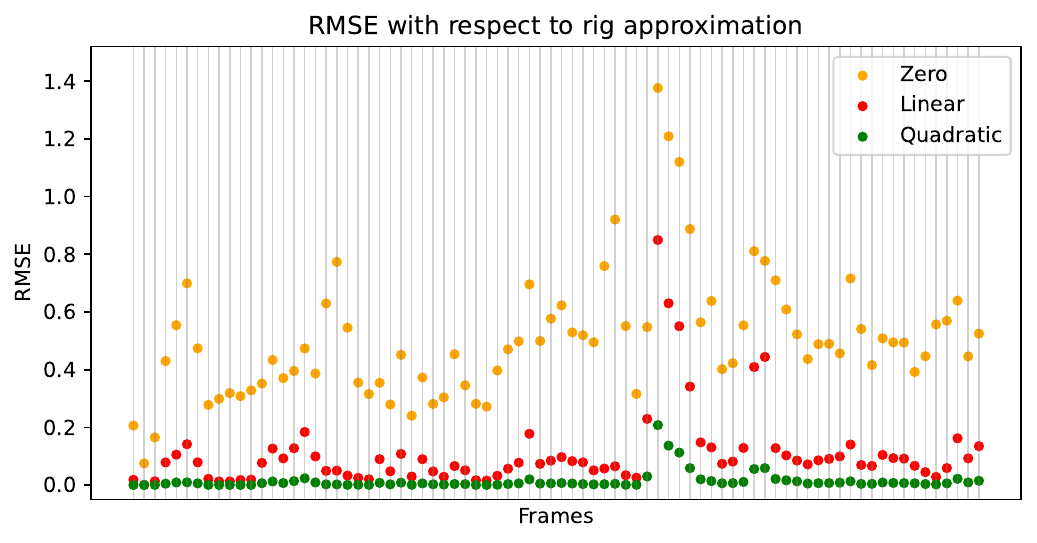}
    \caption{RMSE between ground truth meshes and meshes obtained using different approximations, for the animated character \textit{Omar}. Besides the linear and quadratic blendshape model, an additional \textit{zero} approximation is included to give a better idea of the error scale --- it corresponds to a difference between the original mesh and a neutral expression.}
    \label{fig:rig_approx_err}
\end{figure}

A pair of blendshapes $\textbf{b}_i$ and $\textbf{b}_j$ that deform the same local area can produce mesh artifacts when activated simultaneously, so the additional corrective term $\textbf{b}^{\{i,j\}}\in\mathbb{R}^{3n}$ is included to adjust the resulting deformation. It is constructed as $\textbf{b}^{\{i,j\}} = \widehat{\textbf{b}}^{\{i,j\}} - (\textbf{b}_0+\textbf{b}_i+\textbf{b}_j)$, where $\widehat{\textbf{b}}^{\{i,j\}}$ represents a desired result of joint activation of deformers $i$ and $j$ (an artist sculpts it manually). Now, whenever the blendshapes $\textbf{b}_i$ and $\textbf{b}_j$ are activated simultaneously, the corrective blendshape $\textbf{b}^{\{i,j\}}$ is activated as well, so that the corrective contribution due to simultaneous activation of $\textbf{b}_i$ and $\textbf{b}_j$ equals $w_iw_j\textbf{b}^{\{i,j\}}.$
A \textit{quadratic blendshape function} $f_Q(\cdot): \mathbb{R}^m\rightarrow\mathbb{R}^{3n}$ can now be defined as 
\begin{equation}\label{eq:quadratic_function}
    f_Q(\textbf{w}) = \textbf{b}_0 + \textbf{Bw} + \sum_{(i,j)\in\mathcal{P}}w_iw_j\textbf{b}^{\{i,j\}},
\end{equation}
where $\mathcal{P}$ represents a set of tuples $(i,j)$ such that there is a quadratic corrective term between corresponding blendshapes $\textbf{b}_i$ and $\textbf{b}_j$. In practice, it is common to use additional levels of corrections, yielding even more complex blendshape models --- all the animated characters that we use for the experiments in this paper have at least one more level of corrections. However, the quadratic terms (first level of correction) already give a considerably more accurate approximation of the rig compared to a simple linear model --- in Figure \ref{fig:rig_approx_err} we compare the error of reconstruction of the ground truth animation frames, under linear and quadratic blendshape model approximations, for the animated character \textit{Omar} (for a complete description of datasets see Section \ref{sec:evaluation}). Regardless of their wide use in practice, the corrective terms of the blendshape model are largely overlooked in the literature.

% -------------------------------------------------------------------------------------------------------------------
\subsection{Inverse Rig Problem}\label{ss:inverse_rig}

In this section, we describe the concept that is central to this paper --- the inverse rig problem for facial animation. Further, we cover state-of-the-art solutions before introducing our novel approach in the next section. 

The \textit{inverse rig problem} in automatic motion retargeting is the problem of finding optimal activation parameters $\textbf{w}$ (blendshape weights) to produce a target mesh $\widehat{\textbf{b}}$, which is usually given as a 3D scan of an actor or a set of MoCap markers. It is common to pose the problem in the least squares framework:
\begin{equation}\label{eq:common}
    \minimize_{\textbf{w}}\|f(\textbf{w}) - \widehat{\textbf{b}}\|^2,
\end{equation}
where $f(\textbf{w}):\mathbb{R}^m\rightarrow\mathbb{R}^{3n}$ is a rig function, $\widehat{\textbf{b}}\in\mathbb{R}^{3n}$ is a target mesh, and additional constraints and regularization terms might be included. Regularization terms are added to produce a more stable solution, but might also help to make vector $\textbf{w}$ sparser --- this is desirable because animators usually need to alter the solutions by hand, which gets much harder if a large number of blendshapes are already activated \cite{seol2011artist}.
In \cite{joshi2006learning}, the objective function takes exactly the form of (\ref{eq:common}); in \cite{choe2001analysis, Liu2010LocalizedOF} the constraints are invoked to keep the weights in the $[0,1]$ interval; \cite{lewis2010direct,ccetinaslan2016position} add a regularization term $\|\textbf{w} \|^2$, while \cite{ribera2017facial} uses a $\|\textbf{w} \|_1$ regularization, which is known to enhance sparsity of the solution. Importantly, all the mentioned papers assume the blendshape model is linear.

We take the formulation given in \cite{cetinaslan2020sketching}, as a state-of-the-art approach using a linear blendshape model:
\begin{equation}\label{er:cetinaslan}
    \minimize_{\textbf{w}} \|\textbf{Bw} - \widehat{\textbf{b}} \|^2 + \alpha\|\textbf{w}\|^2,
\end{equation}
where $\alpha\geq0$. The weight constraints are not explicitly included in the optimization, but the values of the resulting vector that are outside of the feasible set are clipped afterward in order to satisfy the model constraints. Note that a neutral face $\textbf{b}_0$ is omitted in the above formulation, hence the target $\widehat{\textbf{b}}$ is also taken as an offset from the neutral face and not an actual mesh. One adjustment to the above approach, given in the same paper, is using a sparse approximation $\textbf{B}^{loc}$ of a matrix $\textbf{B}$, instead of an actual blendshape matrix. This excludes irrelevant blendshape effects in the local regions and leads to a sparser solution and lower computational cost; however, it might affect the accuracy of the reconstructed mesh. 

A different approach is given by \cite{seol2011artist}, where the problem is solved sequentially, for a single blendshape at a time (\textit{step 1} below), and the residual mesh $\widehat{\textbf{b}}$ is updated after each iteration (\textit{step 2}), before proceeding for the next controller:
\begin{equation}\label{eq:seol}
\begin{split}
   \text{step 1:}\quad & w_i  =  \argmin_{w} \| \textbf{b}_iw - \widehat{\textbf{b}}\|^2 \\
   \text{step 2:}\quad & \widehat{\textbf{b}}  \leftarrow \widehat{\textbf{b}} - \textbf{b}_i w_i.
\end{split}
\end{equation}
The order in which weights are optimized is crucial here. The authors suggest sorting them according to the average magnitude of deformation each blendshape produces over a whole face. This method yields a sparse solution and avoids simultaneous activation of mutually exclusive blendshapes \cite{seol2011artist, lewis2014practice}.

To the best of our knowledge, no prior work considered adding the non-linear terms in the objective when solving the inverse rig problem. While working with linear approximation yields a simple and fast solution, it lacks accuracy and fails to closely resemble a source actor. In the next section, we introduce a method that allows the application of the quadratic blendshape function (\ref{eq:quadratic_function}) to produce an accurate and sparse solution for the inverse rig problem. 

% ####################################################################################################################

\section{Proposed Method}\label{sec:proposed_method}

A linear blendshape function is, in general, convenient for solving the inverse rig problems, especially for cartoonish characters whose face proportions differ significantly from a source actor. However, we propose a more suitable model for a specific subdomain of facial animation --- character face models that are sculpted to closely resemble a source actor, which demands higher accuracy compared to what a linear function offers. We consider the quadratic blendshape function (\ref{eq:quadratic_function}). Additionally, we assume that activation weights $w_1,...,w_m$ must stay within the $[0,1]$ interval\footnote{Some authors allow for values outside this interval, either for the sake of simplicity or because it might be beneficial for cartoon characters. However, in the subdomain of animation targeted by our work, the construction of the animation models does not permit violating this constraint.}.

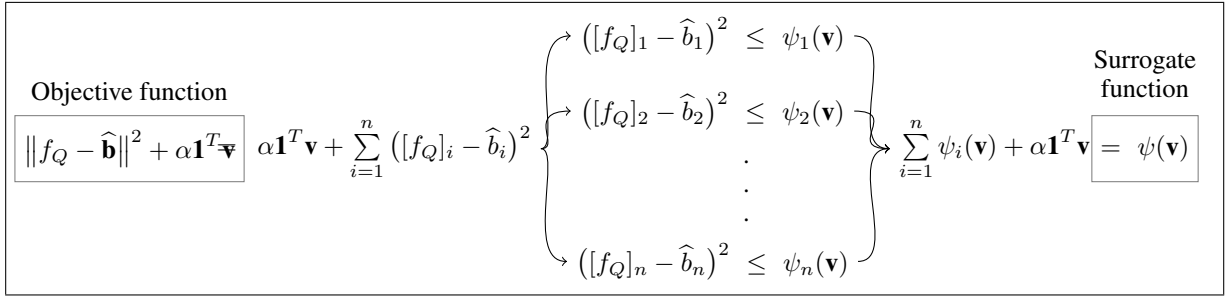
\begin{figure*}
\begin{center}
\frame{
        \begin{tikzpicture}
        \tikzset{LeafVertex/.style = {shape=rectangle,draw=gray,minimum size=2.5em}}
        \tikzset{vertex/.style = {shape=rectangle,draw=none,minimum size=2.5em}}
        \tikzset{edge/.style = {->,> = latex'}}
        \node[vertex] (xxx1) at  (-1.2,1.5) {};
        \node[vertex] (xxx2) at  (14.1,1.5) {};
        \node[LeafVertex,label=Objective function] (OF) at  (0,1.5) {$\big\|f_Q-\widehat{\textbf{b}}\big\|^2+\alpha\textbf{1}^T\textbf{v}$};
        \node[vertex] (OF2) at  (3.25,1.5) {$= \,\,\,\alpha\textbf{1}^T\textbf{v} + \sum\limits_{i=1}^n\big( [f_Q]_i - \widehat{b}_i \big)^2 $};
        \node[vertex] (Split1) at  (7.75,3) {$ \big([f_Q]_1 - \widehat{b}_1 \big)^2 \,\,\leq\,\, \psi_1(\textbf{v})$};
        \node[vertex] (Split2) at  (7.75,2) {$ \big([f_Q]_2 - \widehat{b}_2 \big)^2 \,\,\leq\,\, \psi_2(\textbf{v})$};
        \node[vertex] (SplitDots) at  (8.2,1) {\makecell{.\\.\\.}};
        \node[vertex] (SplitN) at  (7.75,0) {$ \big([f_Q]_n - \widehat{b}_n \big)^2 \,\,\leq\,\, \psi_n(\textbf{v})$};
        \node[vertex] (Concat) at  (11.5,1.5) {$\sum\limits_{i=1}^n\psi_i(\textbf{v}) + \alpha\textbf{1}^T\textbf{v} $};
        \node[LeafVertex, label=\makecell{Surrogate \\function}] (SF) at  (13.5,1.5) {$ = \,\, \psi(\textbf{v})$};
        \draw[->] (OF2) .. controls +(right:2.5cm) and +(left:2.5cm) .. node[above,sloped]{} (Split1);
        \draw[->] (OF2) .. controls +(right:2.5cm) and +(left:2.5cm) .. node[above,sloped]{} (Split2);
        \draw[->] (OF2) .. controls +(right:2.5cm) and +(left:2.5cm) .. node[above,sloped]{} (SplitN);
        \draw[->] (Split1) .. controls +(right:2.5cm) and +(left:2.cm) .. node[above,sloped]{} (Concat);
        \draw[->] (Split2) .. controls +(right:2.5cm) and +(left:2.cm) .. node[above,sloped]{} (Concat);
        \draw[->] (SplitN) .. controls +(right:2.5cm) and +(left:2.cm) .. node[above,sloped]{} (Concat);
    \end{tikzpicture}}
    %\captionsetup{type=figure}
    \caption{Graphical representation of the construction of the proposed upper bound function $\psi(\cdot)$. To simplify the representation, in this figure we use a notion $f_Q$ instead of $f_Q(\textbf{w}+\textbf{v})$.}\label{fig:functionSceheme}
\end{center}%
\end{figure*}

We will first formulate the objective function that should be minimized in order to provide the desired solution and show how it can be solved using standard ad-hoc quadratic solvers. However, we will show experimentally that such a solution is not entirely satisfactory; hence we further introduce an algorithm that we developed for solving this specific optimization problem, which yields a solution with an optimal trade-off between several metrics important for the domain application. 

% -------------------------------------------------------------------------------------------------------------------
\subsection{Objective Function}

Our objective will encode the two most desired properties of the solution to the inverse rig problem: (1) the difference between a reference and reconstructed mesh should be as low as possible, hence we want to incorporate the quadratic corrective terms instead of using only a linear approximation of the blendshape model; (2) the solution vector should exhibit low cardinality, i.e., the number of non-zero elements, since that helps both for the posterior manual modification, and ensuring the stability of the solution; and finally, the weights must stay in the feasible interval $[0,1]$. The proposed objective takes into account all the above assumptions:
\begin{equation}\label{eq:inverse_quadratic_problem}
    \minimize_{\textbf{0}\leq\textbf{w}\leq\textbf{1}} \|f_Q(\textbf{w}) - \widehat{\textbf{b}}\|^2 + \alpha \textbf{1}^T\textbf{w}.
\end{equation}
The regularization term in (\ref{eq:inverse_quadratic_problem}), with $\alpha\geq0$, is invoked to encourage a sparse solution --- note that, as $\textbf{w}\geq\textbf{0}$ in the feasible set, $\textbf{1}^T\textbf{w}$ equals the $L1$ norm of $\textbf{w}$, which is known to be a sparsity-enhancing regularizer \cite{schmidt2005least}. 

We show that quadratic terms increase quality by using standard solvers and increase performance with a dedicated algorithm.
Compared to the approaches described in Section \ref{ss:inverse_rig}, this formulation is a hard-to-solve nonconvex problem and, in particular, does not allow for a closed-form solution. However, it can be readily solved using some standard solvers. Specifically, in this paper, we show the results of applying the method from \cite{byrd1999interior} that is a barrier (interior point) method in which the subproblems are solved approximately by a sequential quadratic programming (SQP) iteration with trust regions \cite{conn2000trust, wright1999numerical}. We decided to use \cite{byrd1999interior} as it corresponds to a common python library implementation of a general-purpose solver that is suitable for our problem -- it handles constrained, nonconvex problems -- while utilizing SQP and interior point ideas\footnote{\url{docs.scipy.org/doc/scipy/reference/generated/scipy.optimize.minimize.html} \\ \url{docs.scipy.org/doc/scipy/reference/optimize.minimize-trustconstr.html}}. This approach yields a high-accuracy fit of the solution, thus demonstrating the advantage of using quadratic terms.

However, this is a general-purpose algorithm, and as expected, the solution obtained by it exhibits other issues. We will show in our experiments that, even though the data fidelity is exceptionally high, the resulting weight vector activates all the elements, yielding an undesirable solution. On top of that, frame-to-frame transitions are far from being smooth. To address all these issues, we propose a novel algorithm for solving the problem (\ref{eq:inverse_quadratic_problem}).

% -------------------------------------------------------------------------------------------------------------------
\subsection{Proposed Algorithm}\label{ss:proposed_algorithm}

The objective in (\ref{eq:inverse_quadratic_problem}) is hard to minimize, hence we resort to the \textit{majorization-minimization} \cite{schifano2010majorization, sun2016majorization} technique to obtain a good-enough approximate solution that clearly improves on the linear model. The idea is to approximate the objective with an iteration over functions $\psi(\cdot)$, which are easier to solve. In the following lines, we will explain how the appropriate function is constructed (Figure \ref{fig:functionSceheme}). Further, we will see that it is separable by coordinates, hence we estimate each blendshape weight independently, solving the set of simple problems. This approach is iterative and suitable for parallelization since the weights are fitted independently. Further, as we will confirm in the numerical experiments, the obtained solution is accurate while keeping the low cardinality (the number of non-zero weights) and smooth transitions over the frames. 

We start by rewriting the objective (\ref{eq:inverse_quadratic_problem}) in a \textit{Levenberg-Marquardt} fashion \cite{Ranganathan2004TheLA}, i.e., we assume some initial weight vector $\textbf{w}$ is available, and we are looking for an increment vector $\textbf{v}$ that would lead to a better solution $\textbf{w} \leftarrow \textbf{w}+\textbf{v}$. Optimization problem is then reformulated as 
\begin{equation}\label{eq:increment_problem}
    \minimize_{-\textbf{w}\leq\textbf{v}\leq\textbf{1-w}} \|f_Q(\textbf{w+v}) - \widehat{\textbf{b}}\|^2 + \alpha \textbf{1}^T\textbf{v}.
\end{equation}
A quadratic term in function $f_Q$ makes this problem too complex to solve exactly, so we apply a majorization-minimization paradigm that solves an approximate problem. We introduce an upper bound function $\psi(\textbf{v};\textbf{w}):\mathbb{R}^m\rightarrow\mathbb{R}$ over the above objective (\ref{eq:increment_problem}), such that $ \|f_Q(\textbf{w+v}) - \widehat{\textbf{b}}\|^2 + \alpha \textbf{1}^T\textbf{v} \leq \psi(\textbf{v};\textbf{w}) $ holds for $ \textbf{0}\leq\textbf{w}+\textbf{v} \leq \textbf{1}$. This bound is easier to minimize than the original objective, and we continue with the problem in the form 
\begin{equation}\label{eq:increment_problem_psi}
    \minimize_{-\textbf{w}\leq\textbf{v}\leq\textbf{1-w}} \psi(\textbf{v};\textbf{w}).
\end{equation}
In other words, for a current solution estimate $\textbf{w}$, we search for increment $\textbf{v}$ to construct the new solution estimate $\textbf{w+v}$ by minimizing a surrogate function $\psi(\textbf{v};\textbf{w})$ in (\ref{eq:increment_problem_psi}) instead of the original cost function in (\ref{eq:increment_problem}). The surrogate function (a global upper bound on the cost function in (\ref{eq:increment_problem})) is carefully constructed such that it represents a good, $\textbf{w}$-dependent approximation of the cost in (\ref{eq:increment_problem}) around the current point $\textbf{w}$, and such that (\ref{eq:increment_problem_psi}) is easy to solve. Further, it needs to respect the following two conditions:\\
\begin{itemize}
    \item for any feasible vector $\textbf{0}\leq\textbf{w}\leq\textbf{1}$, and any increment vector $\textbf{v}$ such that $\textbf{0}\leq\textbf{w}+\textbf{v}\leq\textbf{1}$, the original objective is bounded by the surrogate from above, i.e.,
\begin{equation}
    \|f_Q(\textbf{w+v}) - \widehat{\textbf{b}}\|^2 + \alpha \textbf{1}^T\textbf{v} \leq \psi(\textbf{v};\textbf{w});
\end{equation}
    \item at point $\textbf{v}=\textbf{0}$, the values of the objective and of the majorizer function are equal, that is, 
\begin{equation}
    \|f_Q(\textbf{w}) - \widehat{\textbf{b}}\|^2  = \psi(\textbf{0};\textbf{w}).
\end{equation}
\end{itemize}

In the rest of this paper we will write $\psi(\textbf{v})$ instead of $\psi(\textbf{v};\textbf{w})$ for the sake of simplicity. We will consider a surrogate function such that the regularization term in (\ref{eq:increment_problem}) is kept the same, and we only approximate the data fidelity term. Note that the data fidelity term is a sum of squares over the mesh coordinates. Hence, we will look for the bound that can also be written as a sum, and bound the function at each of the coordinates: 
\begin{equation}
    \psi(\textbf{v}) = \sum_{i=1}^{3n} \psi_i(\textbf{v}) + \alpha \textbf{1}^T\textbf{v},
\end{equation}
where $\psi_i(\cdot):\mathbb{R}^m\rightarrow\mathbb{R}$ is constructed to upper-bound the data fidelity term in the mesh coordinate $i$, i.e., $([f_Q(\textbf{w+v})]_i - \widehat{b}_i)^2 \leq\psi_i(\textbf{v}).$ The notation $[f_Q(\textbf{w+v})]_i$ indicates that we are observing the $i^{th}$ coordinate of the resulting mesh. We want to rewrite this in a canonical quadratic matrix form, hence we introduce a symmetric matrix $\textbf{D}^{(i)}\in\mathbb{R}^{m\times m}$, such that its elements are coordinates of the corrective terms for the corresponding blendshape pairs, $D^{(i)}_{jk} = \frac{1}{2}b_i^{\{j,k\}}.$ Now we have $[f_Q(\textbf{w})]_i = \textbf{B}_i^T\textbf{w} + \textbf{w}^T\textbf{D}^{(i)}\textbf{w}$, so the coordinate-wise surrogate function should satisfy 
\begin{equation}
    \bigg(\textbf{B}_i^T  (\textbf{w}+\textbf{v}) + (\textbf{w}+\textbf{v})^T\textbf{D}^{(i)}(\textbf{w}+\textbf{v}) - \widehat{b}_i\bigg)^2 \leq \psi_i(\textbf{v}).
\end{equation}
Since we are minimizing with respect to the increment vector $\textbf{v}$, the vector $\textbf{w}$ is a constant; hence the left-hand side can be simplified to 
\begin{equation}\label{eq:simplified_fidelity}
    \big( g_i + \textbf{h}_i^T\textbf{v} + \textbf{v}^T\textbf{D}^{(i)}\textbf{v} \big)^2,
\end{equation}
where $g_i = \textbf{B}_i^T\textbf{w} + \textbf{w}^T\textbf{D}^{(i)}\textbf{w} - \widehat{b}_i$, and $\textbf{h}_i = \textbf{B}_i + \textbf{w}^T\textbf{D}^{(i)}$ are the terms that do not depend on $\textbf{v}$. The upper bound function is then derived by bounding the non-linear terms of the above expression separately (a linear term is kept the same). A coordinate-wise bound that satisfies the above conditions is 
\begin{equation}
    \psi_i(\textbf{v}) = g_i + \textbf{h}_i^T\textbf{v} +  g_i\lambda_{M}(\textbf{D}^{(i)},g_i) \|\textbf{v}\|^2 +  2\|\textbf{h}_i\|^2\|\textbf{v}\|^2 + 2m\sigma^2(\textbf{D}^{(i)})\sum_{j=1}^mv_j^4,
\end{equation}
where a function $\lambda_M(\cdot,\cdot):(\mathbb{R}^{m\times m})\rightarrow\mathbb{R}$ is defined as 
\begin{equation}
    \lambda_M(\textbf{D}^{(i)},g_i) :=
    \begin{cases}
      \lambda_{\text{min}}(\textbf{D}^{(i)}), & \text{if } g_i<0,\\
      \lambda_{\text{max}}(\textbf{D}^{(i)}), & \text{if } g_i\geq0,
    \end{cases} 
\end{equation}
and $\lambda_{min}(\textbf{D}^{(i)})$, $\lambda_{max}(\textbf{D}^{(i)})$ stand for the minimum and maximum eigenvalues of the matrix $\textbf{D}^{(i)}$ respectively, and $\sigma(\textbf{D}^{(i)})$ is the largest singular value of the matrix $\textbf{D}^{(i)}$.

A complete bound is reached by summing all coordinate-wise bounds and adding a regularization term:
\begin{equation}\label{eq:upper_bound_fcon}
    \psi(\textbf{v}) = \sum_{i=1}^n \bigg( g_i^2 + 2g_i\sum_{j=1}^mh_{ij}v_j + 2\left( g_i\lambda_M(\textbf{D}^{(i)},g_i) + \|\textbf{h}_i\|^2 \right)\sum_{j=1}^m v_j^2 +  2m\sigma^2(\textbf{D}^{(i)})\sum_{j=1}^mv_j^4\bigg) + \alpha \textbf{1}^T\textbf{v}.
\end{equation}
Figure \ref{fig:functionSceheme} illustrates the above process.

One can see that the upper bound (\ref{eq:upper_bound_fcon}) is separable by components, and that the solution will be available in closed form. If we consider a single blendshape index $j\in\{1,...,m\}$ and regroup the coefficients of the bound function as
\begin{equation}\label{eq:coefficients}
        q  = \sum_{i=1}^{3n}g_ih_{ij} + \alpha, \quad
        r  = \sum_{i=1}^{3n}(g_i\lambda_M(\textbf{D}^{(i)},g_i)+\|\textbf{h}_i\|^2), \quad
        s  = m\sum_{i=1}^{3n}\sigma^2(\textbf{D}^{(i)}),
\end{equation} 
we can write an objective function in the form of a quartic equation without a cubic term:
\begin{equation}\label{eq:increment_vector}
\begin{split}
    \minimize_{v_j} \,\, & qv_j + rv_j^2 + sv_j^4, \\
    \text{s.t.   } & 0\leq w_j+v_j \leq1.
\end{split}
\end{equation}

This procedure is summarized in Algorithm \ref{alg:inner_iteration} and we refer to it as an \textit{inner iteration}. Algorithm \ref{alg:inner_iteration} solves the problem (\ref{eq:increment_vector}) for each component to give a full increment vector $\textbf{v}$, that is, solving (\ref{eq:upper_bound_fcon}) is equivalent to solving (\ref{eq:increment_vector}) for each $j=1,...,m$ independently. In this sense, our approach is somewhat similar to the solution of \cite{seol2011artist}, as given in (\ref{eq:seol}); however, we do not update vector $\textbf{w}$ before all the components are optimized, which helps to avoid the issues with making the right update order, and additionally opens the possibility for a parallel implementation of the procedure.  

\begin{algorithm}
\caption{Inner Iteration}\label{alg:inner_iteration}
\begin{algorithmic}
\Require Blendshape matrix $\textbf{B}\in\mathbb{R}^{3n\times m}$, corrective blendshape matrices $\textbf{D}^{(i)}\in\mathbb{R}^{m\times m}$ for $i=1,...,3n$, target mesh $\widehat{\textbf{b}}\in\mathbb{R}^{3n}$, regularization parameter $\alpha>0$ and weight vector $\textbf{w}\in[0,1]^m$.
\Ensure An optimal increment vector $\hat{\textbf{v}}$ as a solution to (\ref{eq:increment_vector}).
\State Compute coefficients $q,r$ and $s$ from eq. (\ref{eq:coefficients}) and solve for an optimal increment vector $\hat{\textbf{v}}$:
\State $r=  2\sum_{i=1}^{3n}(g_i\lambda_M(\textbf{D}^{(i)},g_i)+\|\textbf{h}_i\|^2)$, 
\State $s=  2m\sum_{i=1}^{3n}\sigma^2(\textbf{D}^{(i)})$, 
\For{$j=1,...,m$} 
    \State $q=  2\sum_{i=1}^{3n} g_ih_{ij} + \alpha$
    \State $\hat{v}_j = \argmin_{v} qv + rv^2 + sv^4 $
    \State \qquad s.t. $-w_j\leq v \leq 1-w_j$
\EndFor
\State \Return $\hat{\textbf{v}}$
\end{algorithmic}
\end{algorithm}

The solution of the inner iteration will depend on the initial weight vector, hence we repeat the procedure in Algorithm 1 multiple times in order to provide an increasingly good estimate $\textbf{w}$ of the solution to (\ref{eq:increment_problem}), as explained in Algorithm \ref{alg:the_algorithm}. After each iteration $t=1,...,T$, we update the weight vector as $\textbf{w}_{(t+1)} =\textbf{w}_{(t)} + \textbf{v}_{(t)}$. An initial vector can be chosen anywhere within the feasible space $\textbf{0}\leq\textbf{w}_{(0)}\leq\textbf{1}$, but in Section \ref{sec:evaluation} we mention strategies for initialization based on domain knowledge, that lead to faster convergence and yield better results. 

Additional details on the construction of the surrogate and a discussion of the algorithm convergence are outside of the scope of this work, but we include them in the companion paper \cite{rackovic2022majorization}.
    
\begin{algorithm}
\caption{Proposed Method}\label{alg:the_algorithm}
\begin{algorithmic}
\Require 
Blendshape matrix $\textbf{B}\in\mathbb{R}^{3n\times m}$, corrective blendshapes $\textbf{b}^{\{i,j\}}\in\mathbb{R}^{3n}$ for $(i,j)\in\mathcal{P}$, target mesh $\widehat{\textbf{b}}\in\mathbb{R}^{3n}$, regularization parameter $\alpha>0$, initial weight vector $\textbf{w}_{(0)}\in[0,1]^m$, maximum number of iterations $T\in\mathbb{N}$ and the tolerance $\epsilon>0$.
\Ensure $\widehat{\textbf{w}}$ - an approximate minimizer of the problem (\ref{eq:inverse_quadratic_problem}).
\State For each vertex $i$ compose a matrix $\textbf{D}^{(i)}\in\mathbb{R}^{m\times m}$ from the corrective terms, and extract singular and eigen values ($\sigma,\lambda_{min}, \lambda_{max}$):
\For {$i = 1,...,3n$}
    \For {$(j,k)\in\mathcal{P}$}
        \State $D^{(i)}_{jk} = D^{(i)}_{kj} = 1/2 b^{\{j,k\}}_i.$
    \EndFor
    \State  $\textbf{D}^{(i)} \rightarrow \lambda_{\text{min}}(\textbf{D}^{(i)})$, $\lambda_{\text{max}}(\textbf{D}^{(i)})$, $\sigma(\textbf{D}^{(i)})$.
\EndFor
\For{$t = 0,...,T$}
    \State Compute optimal increment $\hat{\textbf{v}}$ using Algorithm \ref{alg:inner_iteration} 
    \State Update the weight vector $\textbf{w}_{(t)}$:
    \State $\textbf{w}_{(t+1)} = \textbf{w}_{(t)} + \hat{\textbf{v}}$
    \State Check convergence
    \If{$\left| \psi(\hat{\textbf{v}}) - \psi(\textbf{0}) \right|<\epsilon$} 
        \State $\hat{\textbf{w}} \leftarrow \textbf{w}_{(t+1)}$
        \State \Return $\hat{\textbf{w}}$
    \EndIf
\EndFor
\State $\hat{\textbf{w}} \leftarrow \textbf{w}_{(t+1)}$
\State \Return $\hat{\textbf{w}}$
\end{algorithmic}
\end{algorithm}

% ####################################################################################################################

\section{Evaluation}\label{sec:evaluation}

\begin{figure}
    \centering
    \includegraphics[width=0.4\linewidth]{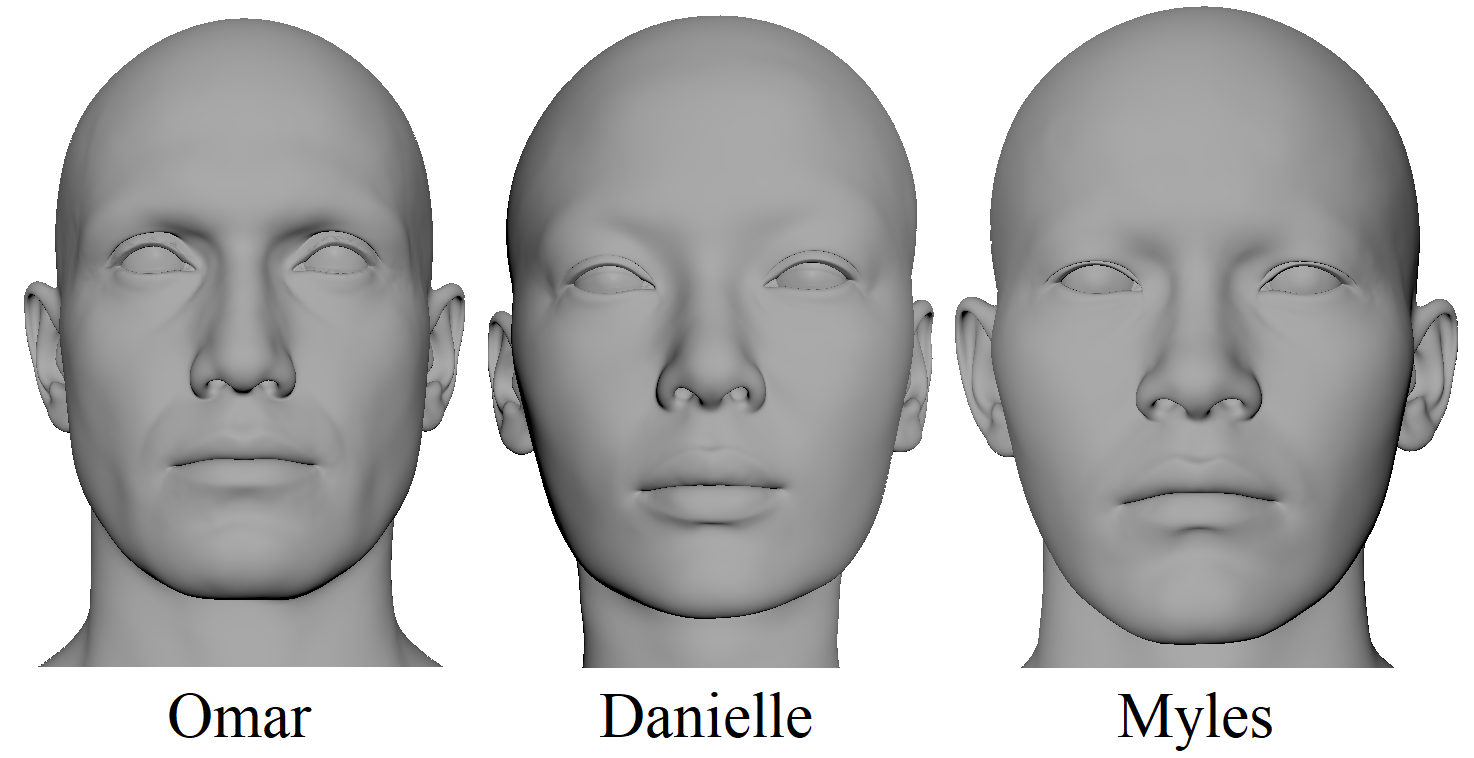}
    \caption{Head models available at MetaHuman Creator platform (\textcopyright unrealengine.com).}
    \label{fig:chars123}
\end{figure}

As mentioned earlier, we consider realistic human characters with a high level of detail and activation weights restricted to lie between 0 and 1. The first three characters that we present in these results are publicly available within the \texttt{MetaHuman}\footnote{We chose \texttt{MetaHuman} characters since they align well with the assumptions of our model and are an increasingly popular choice for building realistic face animation \cite{fang2021metahuman, siddiqui2022fexgan} (\url{unrealengine.com/en-US/metahuman})} platform --- \textit{Omar, Danielle} and \textit{Myles} (\textcopyright unrealengine.com/en-US/eula/mhc), as shown in Figure \ref{fig:chars123}. The additional two datasets that we used to evaluate the method are the property of the animation studio --- \textit{Char 4} and \textit{Char 5}. All the characters are accompanied by a short animation sequence covering a wide range of facial expressions that were used to evaluate the methods\footnote{Upon the acceptance of the paper, the animation weights for \textit{Omar, Danielle} and \textit{Myles} will be available in a public repository, together with the code.}. We exclude inactive vertices and the vertices in the neck and shoulder regions for each character, so after the subsampling, each model has $n=4000$ vertices. The scale of the head is also similar between all the characters, and the width between the left and right ear is approximately $18\, cm$. However, the number of blendshapes differs (ranging between $60$ and $150$), and that means that different choices of the regularization parameter $\alpha\geq0$ (Eq. (\ref{er:cetinaslan}) and (\ref{eq:inverse_quadratic_problem})) might be optimal for various models. It is important to note that \textit{Char 5} has a more complex rig than the other four characters, with a number of deformers that are not based on a blendshape deformation (rotational and joint-like deformers); still, we include it in the experiments to show that our algorithm is robust enough to produce satisfying results even in such a case. 

\subsection{Benchmark Methods}

As mentioned in Section \ref{ss:inverse_rig}, a  state-of-the-art representative of model-based approaches to solving the inverse rig problem is a method given by \cite{cetinaslan2020sketching} in Eq. (\ref{er:cetinaslan}) --- in the rest of this paper we will denote this approach as \textit{Cet}.

In the same paper \cite{cetinaslan2020sketching}, the authors propose a modification of the solution using the heat kernel of \cite{crane2013geodesics} to transform an original blendshape matrix $\textbf{B}$ into a sparse approximation $\textbf{B}^{loc}$. The idea is that the vertices of the face should not affect the activation of blendshapes whose main impact is localized in a distant face region. Hence, this method provides localized and more stable results. The optimization problem for that case is posed identically to (\ref{er:cetinaslan}), except that matrix $\textbf{B}$ is substituted with $\textbf{B}^{loc}$. This approach will be denoted by \textit{Cet-loc}.

A different approach is proposed in \cite{seol2011artist}, where the weights are optimized sequentially (\ref{eq:seol}), starting from the ones that have a larger overall effect on the face. This method does not include any weight constraints, yet our animated characters are strictly demanded to have values between $0$ and $1$. Hence, when applying this method, we project the estimated weight onto a feasible interval after each iteration. This method will be denoted as \textit{Seol}. Note that in this approach, there is no regularization parameter. 

Our method will be evaluated for two cases, as mentioned earlier. The first approach is to consider the proposed objective function (\ref{eq:inverse_quadratic_problem}), but minimize it by applying a general-purpose quadratic solver --- we use the \texttt{scipy} implementation of the interior point solver \cite{byrd1999interior}, whose inner iterations are solved by applying sequential quadratic programming technique and thrust region method\footnote{\url{docs.scipy.org/doc/scipy/reference/optimize.minimize-trustconstr.html} \\ \url{docs.scipy.org/doc/scipy/reference/generated/scipy.optimize.minimize.html}}. We denote this \textit{SQP} solution. The second approach solves the problem (\ref{eq:inverse_quadratic_problem}) using the majorization-minimization-based algorithm proposed in Section  \ref{ss:proposed_algorithm}, and it will be denoted \textit{MM}. The solutions might be affected by choice of the initial point. In Section \ref{sec:proposed_method}, we mentioned that the algorithm might be initialized with any feasible \textbf{w}; however, random initialization leads to slower convergence, and the results are often poor in terms of both mesh fidelity and sparsity, hence we aim for a more educated guess of the initial vector. A simple choice is $\textbf{w}=\textbf{0}$, as this would lead to a sparse solution. The results of our method obtained with zero initialization will be denoted \textit{MM-0}. Another possibility is initializing our method with the solution of a problem under the linear blendshape function approximation. In this case, we chose a solution of \textit{Cet} because it gives a solution in closed form using a pseudoinverse of a matrix $\textbf{B}$; hence, we refer to this as \textit{MM-psd}. In the same way, we can use the results of \textit{Seol} to initialize our method, however, it does not show any advantage over the other, hence we proceed with the above two initialization strategies. (\textit{SQP} is also an iterative solver, but in our experiments different initialization strategies did not show significant differences in the results. For this approach, all the presented results are initialized with a zero vector.)

\subsection{Metrics}\label{ss:metrics}

In order to evaluate the goodness of the results, we will consider several different metrics. The main metric of interest is a root mean squared error (RMSE), which will serve to measure the data fidelity of the results and is defined as:
\begin{equation}\label{eq:mse}
    RMSE(\widehat{\textbf{b}},\tilde{\textbf{b}}) = \sqrt{ \frac{1}{n}\sum_{i=1}^n (\widehat{b}_i - \tilde{b}_i)^2},
\end{equation}
where $\widehat{\textbf{b}}$ is a target face mesh and $\tilde{\textbf{b}}$ is a predicted mesh. The meshes $\tilde{\textbf{b}}$ are obtained by plugging the predicted weight vectors in \texttt{Autodesk Maya} \cite{maya}, hence all the methods are evaluated on the same level of rig complexity. One important thing to note here is that RMSE in this standard form might not be the best choice for assessing the results of mesh reconstruction. The problem is that, when a large number of vertices is only slightly misplaced, the value of the metric might be similar to the case when only a small group of vertices is misplaced but to a large extent. We need to rate these two scenarios differently since the latter is what actually produces visible misfits. In order to accomplish this, we will not use the mean value of the squared distances as in (\ref{eq:mse}), but instead, we take the $95^{th}$ percentile of the error. 

While the mesh error is the metric of primary importance to us, we also want a solution to be as sparse as possible while keeping the mesh fidelity high (i.e., RMSE low). An appropriate metric for this is the cardinality of a predicted weight vector, i.e., the number of non-zero elements of $\textbf{w}$. Some estimated weights might have values that are very close to zero and, in practice, negligible, but they will still count when measuring cardinality. For this reason, we include an additional indicator of the sparsity --- $L_1$ norm of the solution vector. 

Finally, as mentioned earlier, we would prefer if the weights for the consecutive frames of the animation sequence have smooth transitions. If we take a single blendshape weight $w_i$ and observe its values over $T$ time frames, we get a discrete time series $[w_i^{(1)},...,w_i^{(T)}]$. To evaluate its smoothness, we will use the second-order differences \cite{green1993nonparametric, mehlum1998invariant}
\begin{equation}
    \text{Smoothness factor}(w_i) = \sum_{t=2}^{T-1} (w_i^{(t-1)} - 2w_i^{(t)} + w_i^{(t+1)})^2.
\end{equation}
Lower values indicate smoother time series, while the minimum value of $0$ is achieved for a constant vector. Note also that this metric is evaluated per component over time, while the previous ones take a static frame and evaluate over the entire set of weights/vertices. 

\subsection{Numerical Results}

The experiments are executed on a user-level computer with the processor \texttt{Intel(R) Core(TM) i7-8550U CPU @ 1.80GHz   1.99 GHz} and \texttt{8GB} of RAM. 

Our method, under \textit{SQP} and under \textit{MM}, as well as the other benchmark models, are tested with a wide range of regularization parameter values $\alpha\in\{0, 0.001, 0.01, 0.1, 1, 10, 100\}$. We chose this interval of values as it covers both extremely low and extremely high values, hence we expect the optimal value to be somewhere within. However, a reader should be aware that this range might be considerably different depending on the considered animation character. For example,  in \cite{cetinaslan2020sketching}, the authors mention that the values of $\alpha>1$ make the regularization term completely dominates the objective, yet in their experiments with direct manipulation, only a few face vertices are included in the objective. On the other side, we use $n=4000$ vertices for each character in our retargeting experiments; hence, the regularization term takes dominion only when the value of $\alpha$ is much closer to $100$. Besides the number of vertices $n$, the size of the head model (or the units of measure) and the number of blendshapes $m$ will also affect the choice of $\alpha$.

\begin{figure*}
\begin{center}
    \centering
    %\captionsetup{type=figure}
    \textit{Omar \qquad\qquad\qquad\qquad Danielle \qquad\qquad\qquad\qquad Myles \qquad\qquad\qquad\qquad Char 4 \qquad\qquad\qquad\qquad Char 5}
    \includegraphics[width=\linewidth]{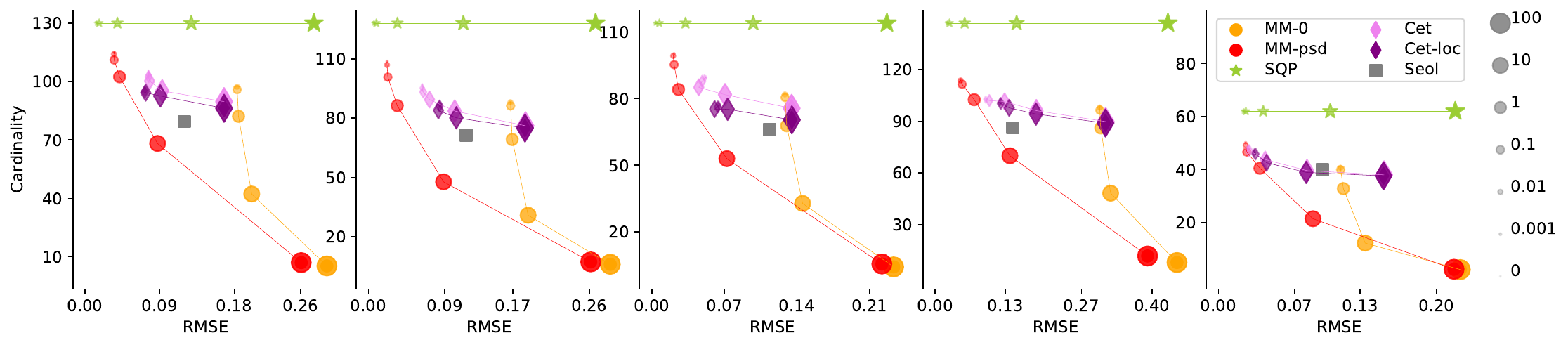}
    \caption{A trade-off between mesh error (RMSE) and cardinality (i.e. the number of non-zero weights) of the estimated solutions for different methods and varying values of regularization parameter $\alpha$. Different marker sizes correspond to values of $\alpha$, as indicated on the far right.}
    \label{fig:tradeoff_card}
    \textit{Omar \qquad\qquad\qquad\qquad Danielle \qquad\qquad\qquad\qquad Myles \qquad\qquad\qquad\qquad Char 4 \qquad\qquad\qquad\qquad Char 5}
    \includegraphics[width=\linewidth]{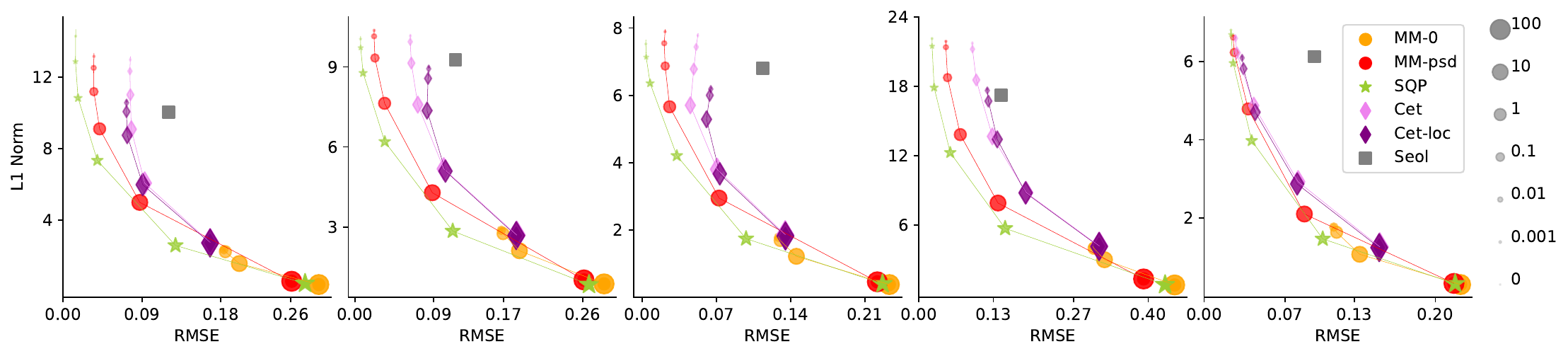}
    \caption{A trade-off between mesh error (RMSE) and L1 norm of the estimated solutions for different methods and varying values of regularization parameter $\alpha$. Different marker sizes correspond to values of $\alpha$, as indicated on the far right.}
    \label{fig:tradeoff_vol}
\end{center}%
\end{figure*}

The desired solution for the rig inversion should have high data fidelity while keeping the number of activated components low, hence we look at the trade-off curves between RMSE (mesh error) and cardinality of the weight vectors in Figure \ref{fig:tradeoff_card}. Additionally, Figure \ref{fig:tradeoff_vol} complements the results showing a trade-off between RMSE and L1 norm of the weight vectors. Predictions are made over the training sets of 80 animated frames for each character. The training frames are subsampled from the original animation sequence so that there are at least 10 time units between any two frames in the resulting set --- this was done in order to avoid the redundancy of the consecutive frames \cite{tena2011interactive}.

The results of \textit{Seol} are presented by a single point (a gray square) in Figures \ref{fig:tradeoff_card} and \ref{fig:tradeoff_vol}, since the approach of \cite{seol2011artist} does not include any regularization. The other methods are presented for different values of $\alpha$ indicated by the size of a corresponding marker and connected by a same-color line. The first thing to notice is that \textit{SQP} (green stars) always activates all the blendshape weights, yielding a trade-off curve that is equal to $m$ on the $y-$axis. On the other side, if we consider the plot of RMSE versus L1 norm, its trade-off is favorable, and actually, \textit{SQP} curves are always under the other methods. This tells us that, most probably, many of those components are activated by a value that is very close to zero, and if further effort is invested, this method might be adjusted to give low cardinality and high-accuracy reconstructions. Results of \textit{MM} are presented in round dots (red for \textit{psd} and orange for 0 initialization). While the zero initialization shows a relatively high mesh error, \textit{MM-psd} gives a trade-off curve that is always below those of the other methods, indicating its superiority. Results of \textit{Cet} and \textit{Cet-loc} (light and dark violet diamonds, respectively) are relatively similar, with the localized version having slightly lower cardinality and higher RMSE. 

To compare the results in more detail, we pick an optimal value of $\alpha$, using the elbow technique, for each approach and each character, and evaluate the predictions over the test set. Test sets consist of 500 frames for each character. In this case, we take the animation sequences instead of the isolated frames. This will allow us to compare the smoothness of the results and to produce the animated clips in order to visually inspect the properties of different methods (check the supplementary video materials).

\begin{figure}
    \centering
    \includegraphics[width=0.6\linewidth]{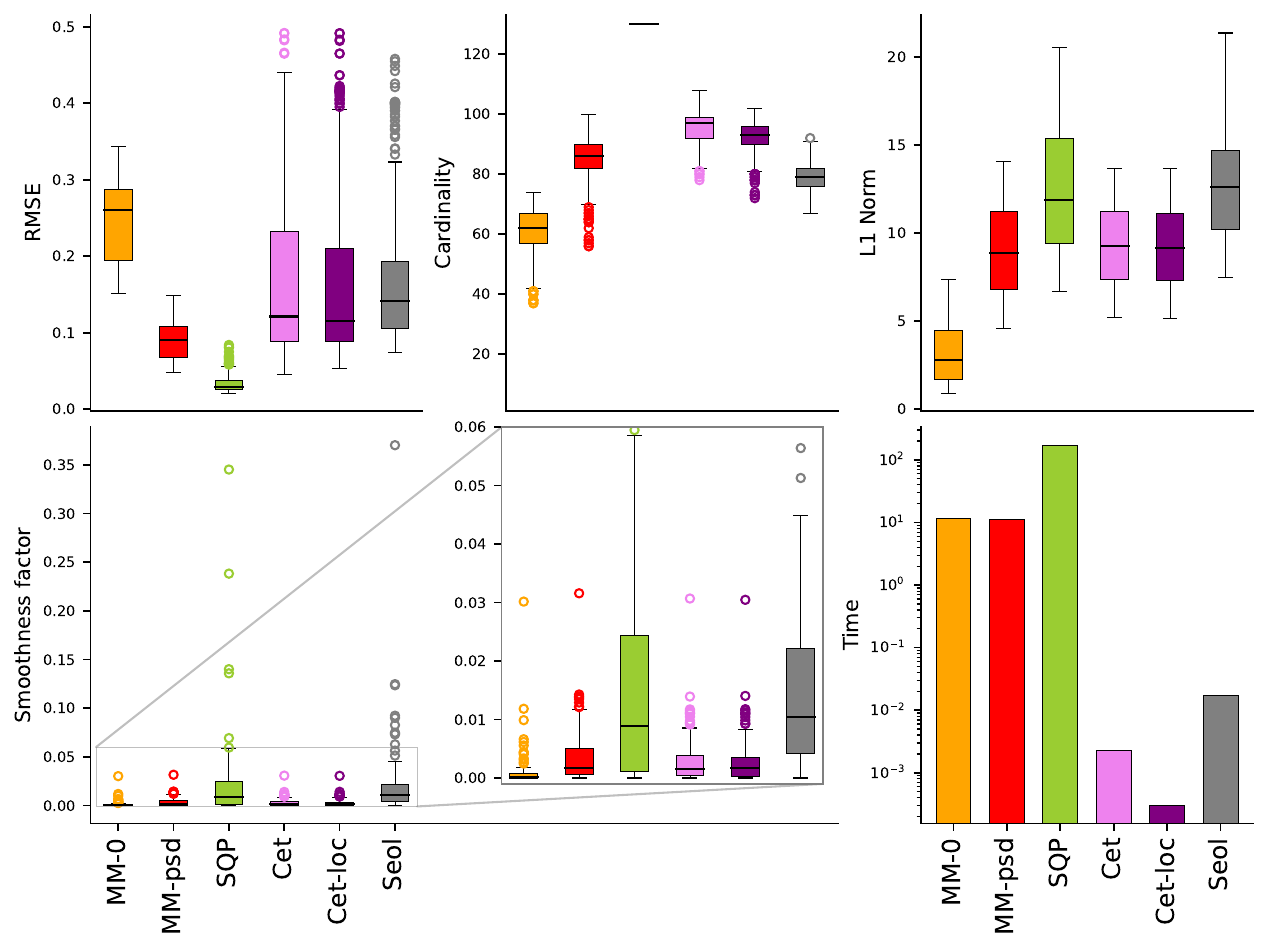}
    \caption{Values of the four metrics (mesh RMSE, weights cardinality, weights $L1$ norm, and temporal smoothness of weight curves) and execution time (in seconds) for \textit{Omar}. Execution time is presented in a log-scale, because of the wide range --- for \textit{Cet-loc} it takes $0.0003$ s, and for \textit{SQP} $172.7$ s. See Table \ref{tab:tabomar} for numerical details.}
    \label{fig:a1_metrics}
\end{figure}

Figure \ref{fig:a1_metrics} gives the values of the four metrics for \textit{Omar} over the test set, as well as the average computational time per frame. Boxplots show the median values and quartiles (across 500 frames in case of RMSE, cardinality, and L1; and across $m=130$ blendshape weights in the case of smoothness factor), while the mean values are given in Table \ref{tab:tabomar}. Table \ref{tab:tabomar} additionally contains the selected optimal $\alpha$ values (notice that the table contains only \textit{MM-psd}, \textit{SQP}, \textit{Cet} and \textit{Seol} as the most significant of the six approaches). Looking at the values of RMSE, \textit{SQP} clearly outperforms all the other approaches; however, it is inferior in all the other respects. Cardinality is, without exception, always set to the maximum value ($m=130$), and the L1 norm is visibly higher than for the other methods, aside from \textit{Seol}. Values of the smoothness factor are far higher than for the other approaches, comparable only to that of \textit{Seol}, hence it was necessary to zoom in on the region of the bottom left subfigure in order to compare the other boxes. Finally, the computation time is much longer than that of \textit{MM}, and since \textit{SQP} is not straightforward to implement in parallel, this cost is irreducible. On the other side, \textit{MM-psd} has a bit higher RMSE than \textit{SQP}, yet still significantly lower than the previous state-of-the-art methods. At the same time, it keeps the cardinality relatively low, somewhere in between \textit{Cet} and \textit{Seol}, and the results are relatively smooth. We address the reader to check the supplementary video materials for a complete visual comparison of the animation sequences.

The only aspect where other methods exhibit better performance is execution time. However, since this method targets the production of movie and game animation, it is not restricted to real-time computations, and the performance speed in our results (around $10 s$ per frame) is feasible. This is especially favorable when contrasted with the solution of a general-purpose solver (\textit{SQP}), which is about 15 times slower. Note that here we implemented the algorithm sequentially, but due to the construction, \textit{inner iterations} could be implemented in parallel to reduce the execution time further.
 
In Figure \ref{fig:teaser}, we see an example frame for character \textit{Omar}, with meshes and activation weights. Again we note that, while the mesh yield by \textit{SQP} gives a flawless reconstruction of the reference mesh, it exhibits too high cardinality of the weights vector, and hence might affect the stability of the result and make  posterior manual editing impossible. On the other side, the three methods under the linear blendshape approximation (\textit{Cet, Cet-loc} and \textit{Seol}) show visible misfits in the mesh --- meshes of \textit{Cet} and \textit{Cet-loc} are very similar, and both are slightly off in the mouth corner region; notice that corners are not as widely spread as in the reference, that the upper lip covers a larger surface of the teeth and that the shadow under the bottom lip is not as visible. \textit{Seol} exhibits low cardinality and lower average error than \textit{Cet}, yet it gives a completely different mouth expression. Our method \textit{MM} is the only one that gives both a good reconstruction of the mesh and low cardinality of the weight vector.

\begin{table}[]
    \centering
    \begin{tabular}{c | c | c | c | c | c |c}
                & RMSE           & Card.   & L1      &  \makecell{Smooth. \\ factor}    & Time         & $\alpha$   \\
                \hline
        MM      & 0.0898        & 85.2           &\textbf{9.10}&  0.0034              &  11.29       &  5         \\
        SQP     &\textbf{0.0325}& 130.           & 12.4        &  0.0201              &  172.7       &  5       \\
        Cet     & 0.1671        & 95.4           & 9.38        &  \textbf{0.0029}     &\textbf{0.002}&  5        \\
        Seol    & 0.1625        & \textbf{78.9}  & 12.7        &  0.0207              &  0.017       &  /
    \end{tabular}
    \caption{Comparison of the obtained metrics values for the four selected approaches, for \textit{Omar}, where $MM$ is with $psd$ initialization. The best value for each metric is given in bold. The last column gives an optimal value of a regularization parameter $\alpha$.}
    \label{tab:tabomar}
\end{table}

\begin{figure}
    \centering
    \includegraphics[width=0.6\linewidth]{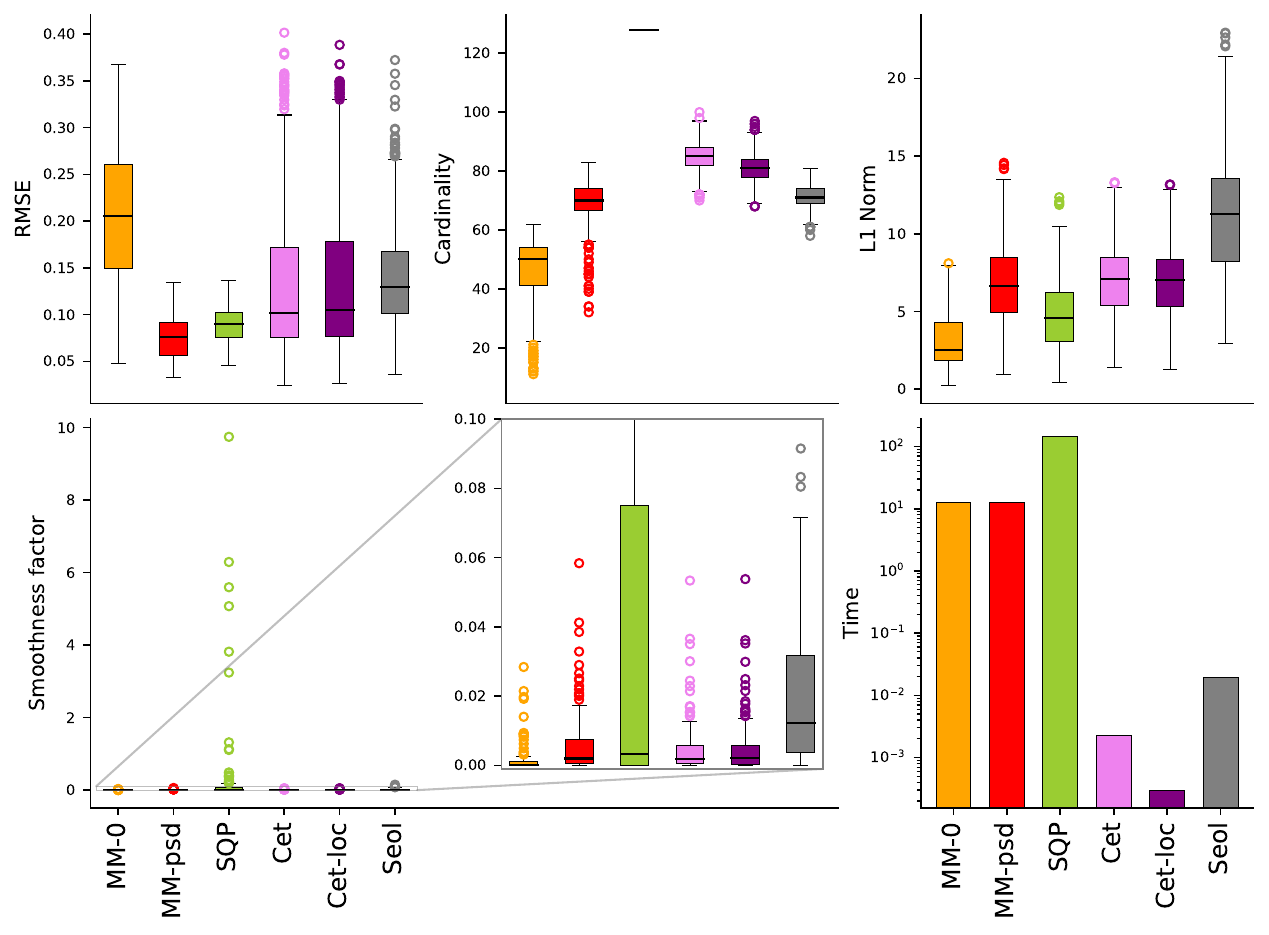}
    \caption{Values of the three metrics (mesh RMSE, weights cardinality, weights $L1$ norm, and temporal smoothness of weight curves) and execution time (in seconds) for \textit{Danielle}. Execution time is presented in a log-scale, because of the wide range --- for \textit{Cet-loc} it takes $0.0003$ s, and for \textit{SQP} $143.2$ s. See Table \ref{tab:tabdanielle} for numerical details.}
    \label{fig:a18_metrics}
\end{figure}

\begin{figure}
    \centering
    \begin{tikzpicture}
    \node[above right, inner sep=0] (image2) at (0.15,2){\includegraphics[width=0.5\linewidth]{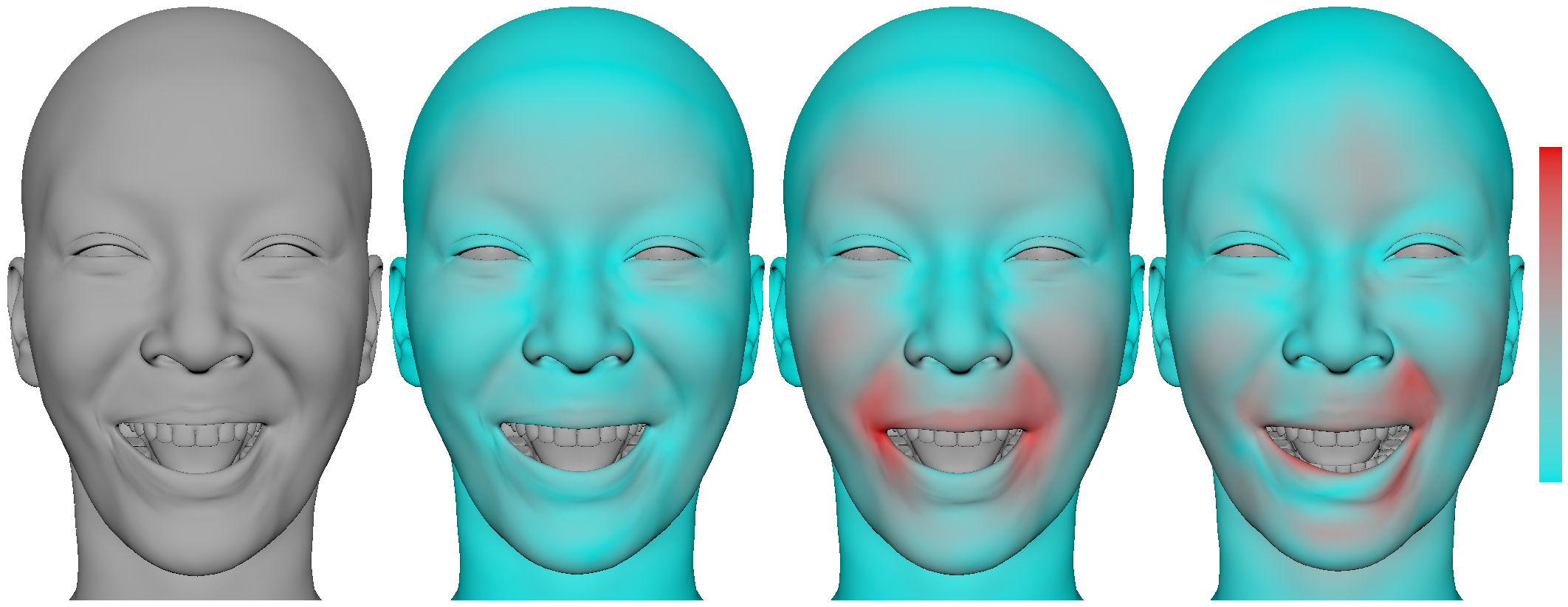}};
    \node[above right, inner sep=0] (image) at (0,0){\includegraphics[width=0.5\linewidth]{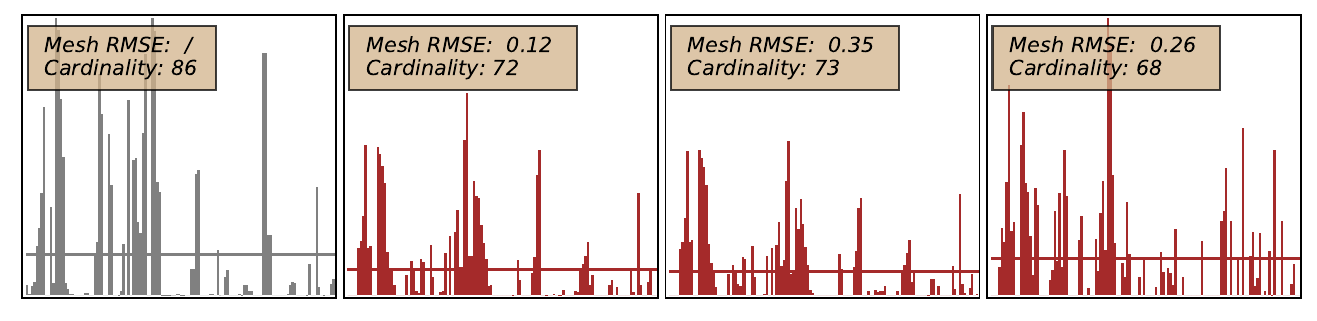}};
    \begin{scope}[
        x={($0.1*(image.south east)$)},
        y={($0.1*(image.north west)$)}]
      \node[darkgray] at (1.30,0.0) {\small Reference };
      \node[darkgray] at (3.80,0.0) {\small MM (ours) };
      \node[darkgray] at (6.30,0.0) {\small Cet };
      \node[darkgray] at (8.80,0.0) {\small Seol };
            \node[darkgray] at  (10,23){\footnotesize .45 };
            \node[darkgray] at  (10,12){\footnotesize .00 };
            \node[darkgray] at  (10,11){\footnotesize cm };
    \end{scope}
    \end{tikzpicture}
    \caption{Example frame prediction for \textit{Danielle}. The top row shows obtained meshes, while the bottom represents corresponding activations of the controller weights. Red tones in the meshes indicate a higher error of the fit, according to the color bar on the right. The average weight activation of each solution is indicated with a horizontal line. The average mesh error and cardinality (i.e. the number of non-zero weights) of the solution are given in a text box.}
    \label{fig:a18_meshes}
\end{figure}
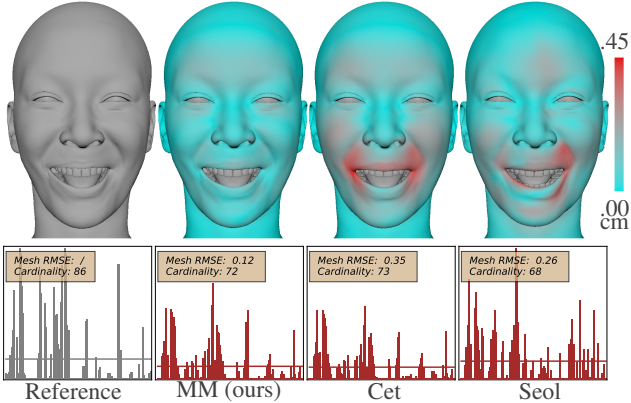

\begin{table}[]
    \centering
    \begin{tabular}{c | c | c | c | c | c |c}
                & RMSE           & Card.         & L1           &  \makecell{Smooth. \\ factor} & Time & $\alpha$ \\
                \hline
        MM      & \textbf{0.0756}& \textbf{69.3} & 6.72         &  0.0062              & 12.73         &  5         \\
        SQP     & 0.0894         & 128.          & \textbf{4.64}&  0.3390              & 143.2         &  5         \\
        Cet     & 0.1280         & 84.6          & 7.02         &  \textbf{0.0051}     &\textbf{0.002} &  5         \\
        Seol    & 0.1389         & 71.5          & 11.1         &  0.0212              & 0.019         &  /
    \end{tabular}
    \caption{Comparison of the obtained metrics values for the four selected approaches, for \textit{Danielle}. The best value for each metric is given in bold. The last column gives an optimal value of a regularization parameter $\alpha$.}
    \label{tab:tabdanielle}
\end{table}

\begin{table}[]
    \centering
    \begin{tabular}{c | c | c | c | c | c |c}
                & RMSE          & Card.       & L1          &  \makecell{Smooth. \\ factor} & Time   & $\alpha$\\
                \hline
        MM      &\textbf{0.0556}&  69.0       & 4.48        & 0.0027                &  8.276         & 5       \\
        SQP     &0.0809         &  114.       &\textbf{2.66}& 0.0118                &  164.3         & 5       \\
        Cet     & 0.0682        &  80.9       & 4.90        & \textbf{0.0024}       &  \textbf{0.004}& 5     \\
        Seol    & 0.1140        &\textbf{65.4}& 7.47        & 0.0181                &  0.015         &  /
    \end{tabular}
    \caption{Comparison of the obtained metrics values for the four selected approaches, for \textit{Myles}. The best value for each metric is given in bold. The last column gives an optimal value of a regularization parameter $\alpha$.}
    \label{tab:tabmyles}
\end{table}

Metric values for \textit{Danielle} are given in Figure \ref{fig:a18_metrics} and Table \ref{tab:tabdanielle}. The conclusions are somewhat similar to the case of \textit{Omar}. Here, \textit{MM} actually gives lower RMSE than \textit{SQP} (and any other approach), and even the cardinality is lower than that of \textit{Seol}. Another notable difference is that \textit{SQP} contrast even more with the other methods in term of smoothness, completely dominating the corresponding subfigure. The example frame with predictions for \textit{MM-psd}, \textit{Cet}, and \textit{Seol} is presented in Figure \ref{fig:a18_meshes}. Both \textit{Cet} and \textit{Seol} show higher errors in the region around the mouth (as indicated by the red color), with \textit{Cet} giving slightly less stretched mouth corners compared to the original, while \textit{Seol} poorly reconstructs the lower lip --- the bottom row of teeth is completely visible, while it is occluded in the reference frame. Our method gives a mesh reconstruction that closely resembles the original, while the cardinality is comparable to the benchmark methods.

\begin{figure}
    \centering
    \includegraphics[width=0.6\linewidth]{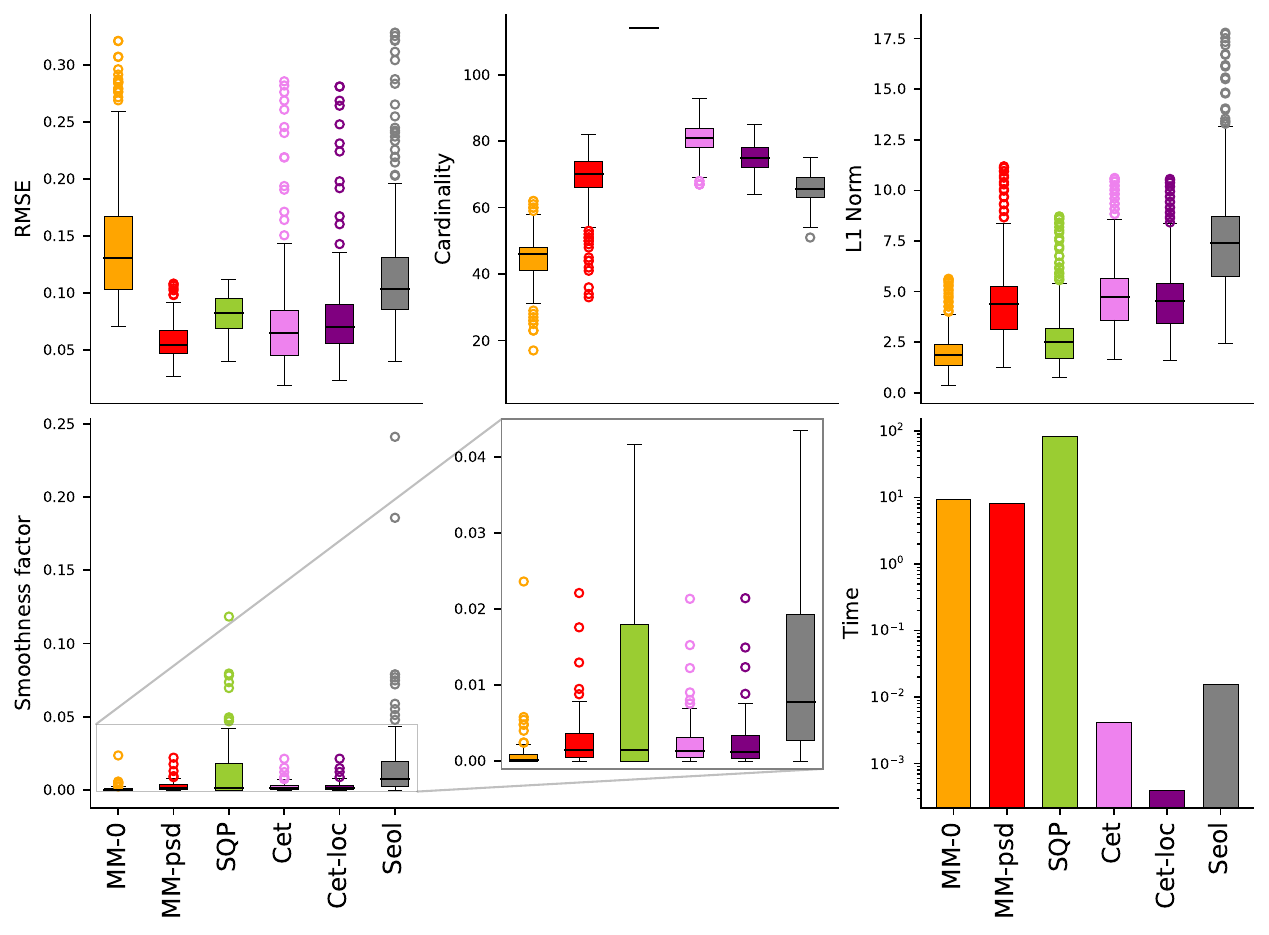}
    \caption{Values of the four metrics (mesh RMSE, weights cardinality, weights $L1$ norm, and temporal smoothness of weight curves) and execution time (in seconds) for \textit{Myles}. Execution time is presented in a log-scale, because of the wide range --- for \textit{Cet-loc} it takes $0.0004$ s, and for \textit{SQP} $164.3$ s. See Table \ref{tab:tabmyles} for numerical details.}
    \label{fig:a21_metrics}
\end{figure}

\begin{figure}
\centering
    \begin{tikzpicture}
    \node[above right, inner sep=0] (image2) at (.15,2.1){\includegraphics[width=.5\linewidth]{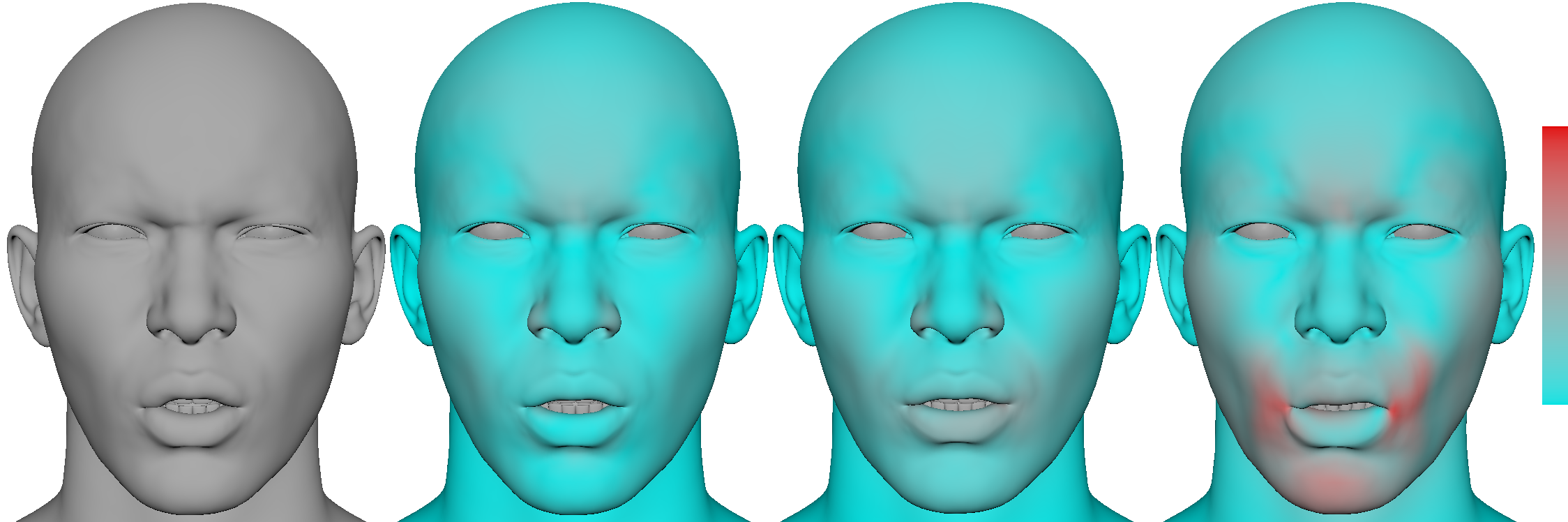}};
    \node[above right, inner sep=0] (image) at (0,0){\includegraphics[width=0.5\linewidth]{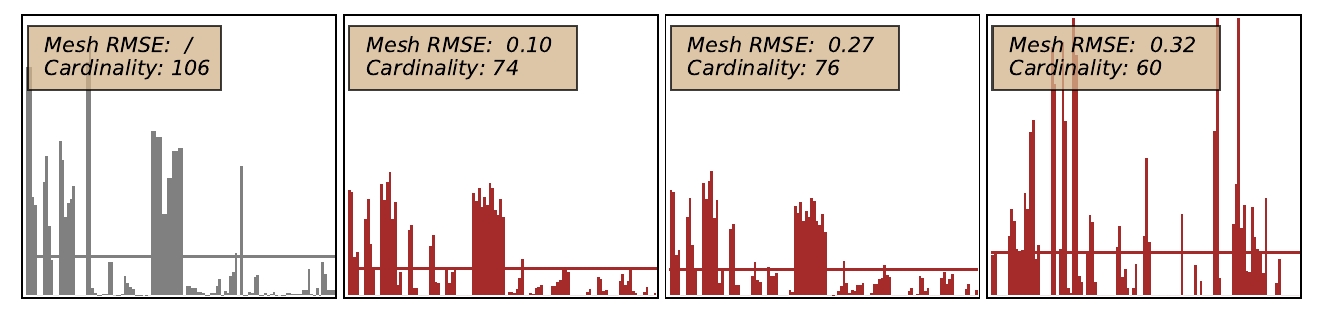}};
    \begin{scope}[
        x={($0.1*(image.south east)$)},
        y={($0.1*(image.north west)$)}]
      \node[darkgray] at (1.30,0.0) {\small Reference };
      \node[darkgray] at (3.80,0.0) {\small MM (ours) };
      \node[darkgray] at (6.30,0.0) {\small Cet };
      \node[darkgray] at (8.80,0.0) {\small Seol };
            \node[darkgray] at  (10.1,22){\footnotesize .49 };
            \node[darkgray] at  (10.1,12.5){\footnotesize .00 };
            \node[darkgray] at  (10.1,11.1){\footnotesize cm };
    \end{scope}
    \end{tikzpicture}
    \caption{Example frame prediction for \textit{Myles}. The top row shows obtained meshes, while the bottom represents corresponding activations of the controller weights. Red tones in the meshes indicate a higher error of the fit, according to the color bar on the right. The average weight activation of each solution is indicated with a horizontal line. The average mesh error and cardinality of the solution are given in a text box.}
    \label{fig:a21_meshes}
\end{figure}
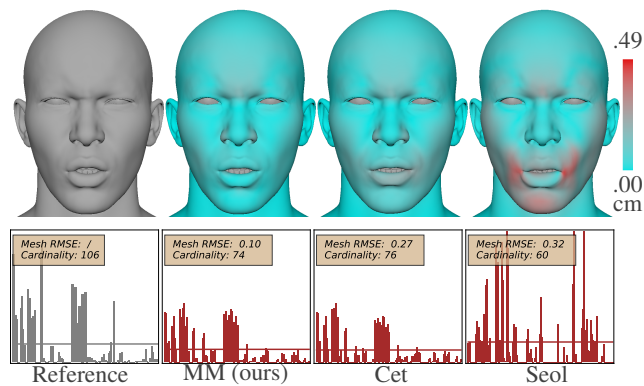

For \textit{Myles}, boxplots are presented in Figure \ref{fig:a21_metrics} (and mean values in Table \ref{tab:tabmyles}). Our method \textit{MM} gives the lowest RMSE and the cardinality that is in between that of \textit{Cet} and \textit{Seol}. In this case, \textit{Cet} (and \textit{Cet-loc}) gives quite low RMSE --- in the median value, it is even lower than \textit{SQP}, yet it varies much more, giving significantly larger upper quantiles. The example frame for \textit{Myles} is in Figure  \ref{fig:a21_meshes}. While the average RMSE is larger (and more red tones are visible in the mesh) for \textit{Cet} than for \textit{MM}, neither of the two produces visible flaws, and bar plots of the wight activations are very similar. On the other side, \textit{Seol} shows a poor fit in the shapes of the lips and cheeks.

\begin{figure}
    \centering
    \includegraphics[width=0.6\linewidth]{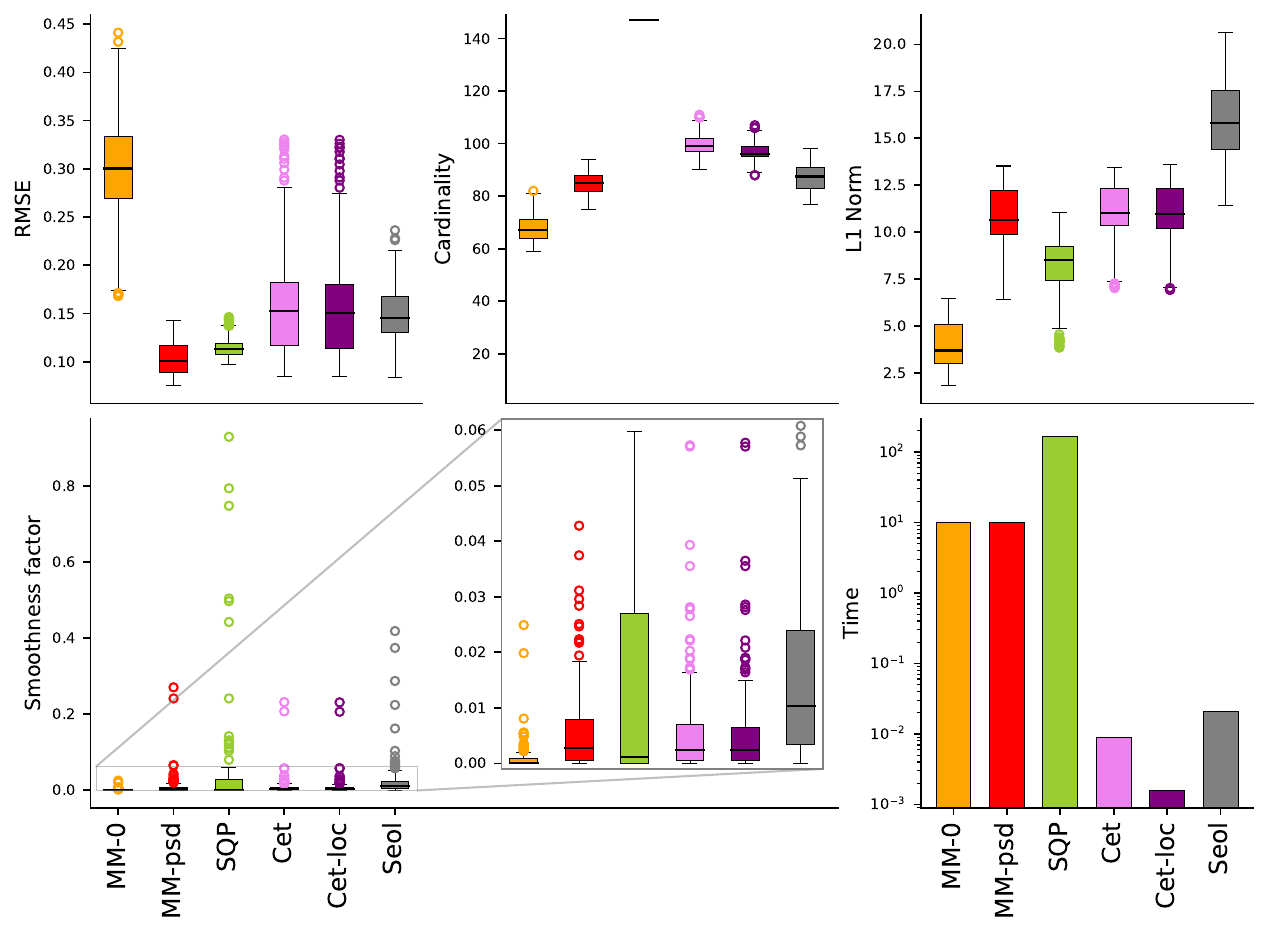}
    \caption{Values of the four metrics (mesh RMSE, weights cardinality, weights $L1$ norm, and temporal smoothness of the weight curves) and execution time (in seconds) for \textit{Char 4}. Execution time is presented in a log-scale, because of the wide range --- for \textit{Cet-loc} it takes $0.0016$ s, and for \textit{SQP} $84.5$ s. See Table \ref{tab:tabC4} for numerical details.}
    \label{fig:hm_metrics}
\end{figure}

\begin{figure}
    \centering
    \includegraphics[width=0.6\linewidth]{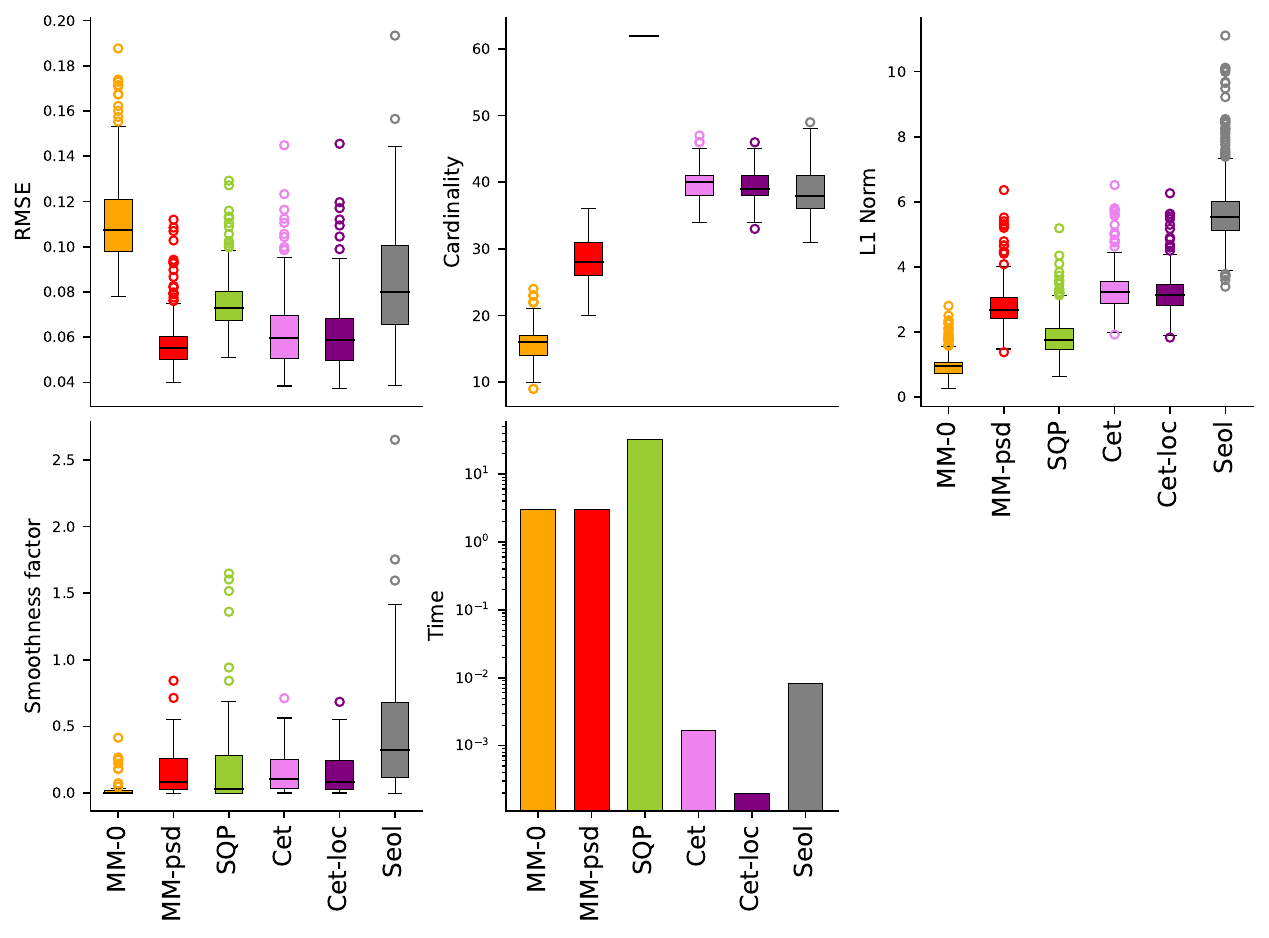}
    \caption{Values of the four metrics (mesh RMSE, weights cardinality, weights $L1$ norm, and temporal smoothness of the weight curves) and execution time (in seconds) for \textit{Char 5}. Execution time is presented in a log-scale, because of the wide range --- for \textit{Cet-loc} it takes $0.0002$ s, and for \textit{SQP} $32.24$ s. See Table \ref{tab:tabC5} for numerical details.}
    \label{fig:o_metrics}
\end{figure}

\textit{Char 4} and \textit{Char 5} are proprietary models, hence we are not showing the meshes; however, the results are summarized in Figures \ref{fig:hm_metrics} and \ref{fig:o_metrics} and in Tables \ref{tab:tabC4} and \ref{tab:tabC5}, respectively. As mentioned earlier, a face model of \textit{Char 5} has a different structure, with many non-blendshape components, while our algorithm targets detailed and accurate blendshape rigs. However, these results show that even with relaxed assumptions about the face model, our method is comparable to state-of-the-art solutions. For \textit{Char 4} the conclusions are similar to those for \textit{Myles}, while for \textit{Char 5} the main takeaway is that \textit{MM} is slightly better than other methods in terms of RMSE, and significantly outperforms the benchmarks in terms of cardinality. A difference in the smoothness factor between the methods is not so drastic here, with only \textit{Seol} showing considerably higher values.

\begin{table}[]
    \centering
    \begin{tabular}{c | c | c | c | c | c |c}
             & RMSE          & Card.       & L1  &  \makecell{Smooth. \\ factor} & Time   & $\alpha$\\
                \hline
        MM   &\textbf{0.1031}&\textbf{85.0}& 10.5        &0.0098         &  10.01       &   5 \\
        SQP  & 0.1142        &  147.       &\textbf{7.92}& 0.0444        &  84.5       &   5 \\
        Cet  & 0.1606        &  99.5       &  10.8       &\textbf{0.0088}&\textbf{0.009}&   5 \\
        Seol & 0.1440        &  87.1       &  15.8       & 0.0263        &  0.021       &   /
    \end{tabular}
    \caption{Comparison of the obtained metrics values for the four selected approaches, for \textit{Char 4}. The best value for each metric is given in bold. The last column gives an optimal value of a regularization parameter $\alpha$.}
    \label{tab:tabC4}
\end{table}

\begin{table}[]
    \centering
    \begin{tabular}{c | c | c | c | c | c |c}
             & RMSE          & Card.       & L1          &  \makecell{Smooth. \\ factor} & Time   & $\alpha$\\
                \hline
        MM   &\textbf{0.0564}&\textbf{28.3}& 2.74        &  0.1638       &  2.985       &  5  \\
        SQP  & 0.0745        &   62.0      &\textbf{1.82}&  0.2189       &  32.24       &  5 \\
        Cet  & 0.0611        &   39.8      &  3.26       &\textbf{0.1550}&\textbf{0.002}&  5   \\
        Seol & 0.0834        &\textbf{38.2}&  5.68       &  0.4908       &  0.008       &  /
    \end{tabular}
    \caption{Comparison of the obtained metrics values for the four selected approaches, for \textit{Char 5}. The best value for each metric is given in bold. The last column gives an optimal value of a regularization parameter $\alpha$.}
    \label{tab:tabC5}
\end{table}

It is interesting to point out that, even though no temporal prior was included in the objective, \textit{MM} and \textit{Cet} yield smooth results for all five datasets, as indicated by the respective smoothness factor and also confirmed by the supplementary video materials. This was not the case for \textit{SQP} or \textit{Seol}. In \cite{seol2011artist}, the solution of \textit{Seol} was modified afterward, using the graph simplification technique, to produce smooth animation curves. However, this was outside of the scope of our paper, hence we did not apply those additional steps. 

\subsection{Animator Feedback}

To make the analysis of the results complete, we include here also the feedback from the expert animators. Five animators were asked to rank the methods (\textit{MM, SQP, Cet, Cet-loc, Seol}) from the best performing to the worst. The criterion was the appearance of the animated results compared to the reference motion --- the animators were looking at the flip-tests provided as the supplementary video materials, except that the method names were masked to avoid possible bias. 

The answers of the animators were heterogenous: sometimes the same animator would give different rankings of the methods for different animated characters (\textit{Omar, Danielle, Myles}), sometimes there would be ties between two or more methods, and in some cases, the animator would only be able to tell the best and the worst of five results. This leaves us with a set of pairwise comparisons of the results. We will estimate the overall ranking of the five methods using the \textit{Bradley-Terry model} \cite{bradley1952rank, chen2013pairwise}, as it is a common ranking procedure when the outcomes of the pairwise comparisons are available.  

\begin{table}[H]
 \centering
\begin{tabular}{ c | c c c c c }
         & \footnotesize MM & \footnotesize SQP & \footnotesize Cet & \footnotesize Cet-loc & \footnotesize Seol \\ 
 \hline 
 \footnotesize MM      & -  & 5   & 12  & 12      & 11 \\  
 \footnotesize SQP     & 6  & -   & 9   & 9       & 12 \\  
 \footnotesize Cet     & 0  & 3   & -   & 3       & 3  \\ 
 \footnotesize Cet-loc & 0  & 2   &  2  & -       & 3  \\ 
 \footnotesize Seol    & 2  & 3   & 5   & 5       & -
\end{tabular}
\caption{Matrix $\textbf{S}$ of the outcomes of pairwise comparisons between the five methods. A value in row $i$ and column $j$ tells how many times the $i^{th}$ method ranked better than the $j^{th}$.}
    \label{tab:tab_bradley}
 \end{table}
 
The Bradley-Terry model is an iterative procedure for estimating the relative strength of the objects (methods in our case) from the set of pairwise comparisons. The outcomes of the comparisons are collected into a square matrix $\textbf{S}\in\mathbb{R}^{N\times N}$, as seen in Table \ref{tab:tab_bradley}; the value in row $i$ and column $j$ shows how many times the $i^{th}$ object (method) ranked better than the $j^{th}$, while the value in a row $j$ and column $i$ indicates the number of the opposite outcomes. From Table \ref{tab:tab_bradley}, we can already see that the values in the first row are the highest, telling us that \textit{MM} was most often ranked as the better method compared to the others, while \textit{SQP} (the second row) is slightly behind. Also, we see that the values of the first column are relatively lower than for the others, confirming that \textit{MM} rarely lost the pairwise comparisons. 

\begin{figure}
    \centering
    \includegraphics[height=.2\linewidth]{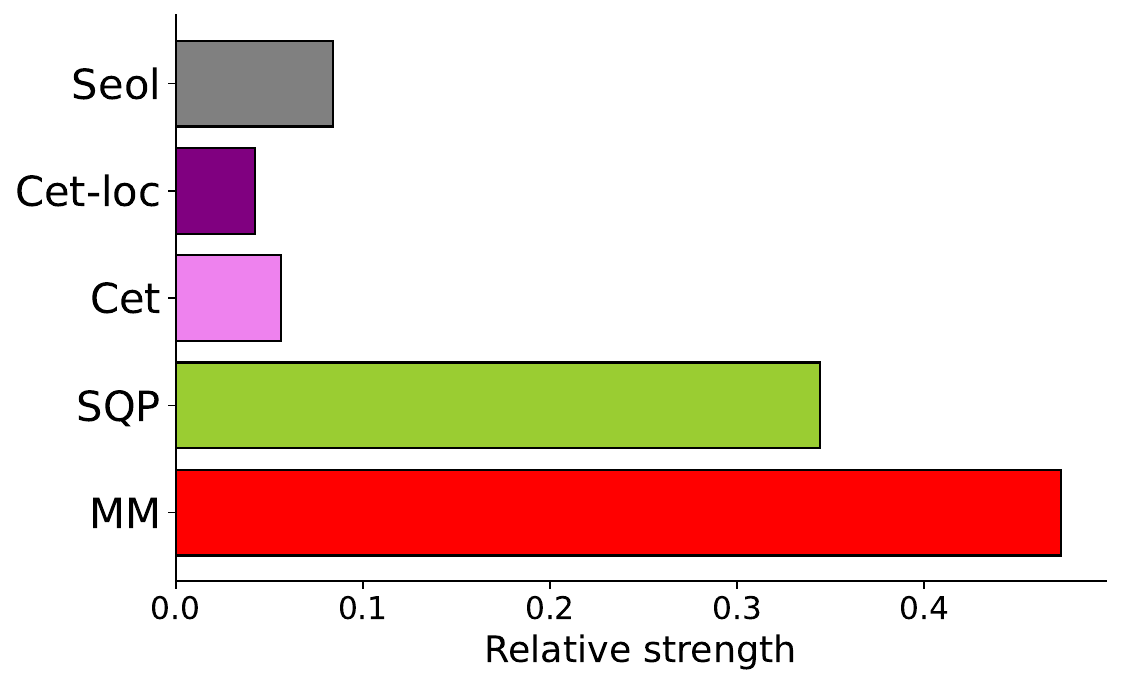}
    \caption{Bradley-Terry estimate of the relative strength vector $\textbf{s}$. Methods that use corrective terms (\textit{MM, SQP}) show a clear superiority over simpler methods.}
    \label{fig:bradley}
\end{figure}

Further, the vector of relative strengths of the methods, $\textbf{s}\in\mathbb{R}^N$, is estimated iteratively, as follows. For each object $i$, the $i^{th}$ coordinate of the vector is 
\begin{equation}
    s_i \leftarrow \frac{\sum_{j\neq i}S_{ij}}{\sum_{j\neq i}\frac{S_{ij}+S_{ji}}{s_i+s_j}}.
\end{equation}
The final values of the relative strength vector $\textbf{s}$, scaled so that it sums to 1, are presented in Figure \ref{fig:bradley}. \textit{MM} shows a clear advantage over the other methods, and it is followed by \textit{SQP}. Other methods are far behind these two. While \textit{SQP} was often showing a lower RMSE than MM, the ranking was more in favor of \textit{MM}, and we can interpret this by the fact that \textit{SQP} produces high cardinality, which led to less stable (or less smooth) transitions between the frames of animation; hence, the estimated animation sequences did not look as organic as \textit{MM}, leading to the lower relative score assigned by the animators. 

Some remarks that animators made, aside from the rankings, are that the differences are easiest to tell in the region of the mouth, especially when the corners are wide open, and that \textit{MM} and \textit{SQP} show a closer fit in this region. This is something that we could have expected since the two methods include the corrective terms when fitting, and the majority of the quadratic corrective terms are usually targeting the lips of the characters. 

% ####################################################################################################################

\section{Conclusion}

The method proposed in this paper applies the majorization minimization paradigm in order to easily achieve the solution to the problem of rig inversion even when the quadratic blendshape terms are utilized. Our method gives a better fit in the details of the face mesh while not increasing the cardinality of the weight vector compared to state-of-the-art methods proposed in \cite{seol2011artist} and \cite{cetinaslan2020sketching}; hence, the method is highly applicable in realistic face animation and targets the applications where accuracy is preferable to real-time execution such as the close shot animations in video games or movies production. It is further worth mentioning that the construction of the algorithm gives space for the parallel implementation of the inner iterations, and in future work, we will address this to reduce the execution time additionally. Another aspect that we will address in future research is to include an additional step of face segmentation. This might lead to distributed model and possibly even higher precision in fitting the fine details of the face mesh. 

% ####################################################################################################################

\textbf{Video Materials}
Flip-tests for each of the three MetaHumans are available at:
\begin{itemize}
    \item Omar \url{https://youtu.be/7RJo9KLaM48} for the colored version and \url{https://youtu.be/AVztr9sOBhY} for the gray version.
    \item Danielle \url{https://youtu.be/yjMS8D1He20} colored and  \url{https://youtu.be/mhAHGYeA-fY} gray.
    \item Myles \url{https://youtu.be/Auy3vE1J8r0} colored and \url{https://youtu.be/r2MraUrX1ew} gray.
\end{itemize}

\textbf{Acknowledgements}

The authors would like to thank 3Lateral studio for guidelines in the animation applications domain, and in specific a group of animators, Aleksa Bračić, Nikola Stošić, Đorđe Ilić, Igor Erić, and Lazar Damjanov, for the valuable feedback on the animated results. We want to thank Dr. Filipa Valdeira for sharing her expertise in the ranking methods to help us evaluate the animators' feedback.

\textbf{Funding}

This work has received funding from the European Union's Horizon 2020 research and innovation program under the Marie Skłodowska-Curie grant agreement No. 812912, from FCT IP strategic project NOVA LINCS (FCT UIDB/04516/2020) and project DSAIPA/AI/0087/2018. The work has also been supported in part by the Ministry of Education, Science and Technological Development of the Republic of Serbia (Grant No. 451-03-9/2021-14/200125).

\bibliographystyle{unsrtnat}
\bibliography{references}  %%% Uncomment this line and comment out the ``thebibliography'' section below to use the external .bib file (using bibtex) .

%%% Uncomment this section and comment out the \bibliography{references} line above to use inline references.
% \begin{thebibliography}{1}

% 	\bibitem{kour2014real}
% 	George Kour and Raid Saabne.
% 	\newblock Real-time segmentation of on-line handwritten arabic script.
% 	\newblock In {\em Frontiers in Handwriting Recognition (ICFHR), 2014 14th
% 			International Conference on}, pages 417--422. IEEE, 2014.

% 	\bibitem{kour2014fast}
% 	George Kour and Raid Saabne.
% 	\newblock Fast classification of handwritten on-line arabic characters.
% 	\newblock In {\em Soft Computing and Pattern Recognition (SoCPaR), 2014 6th
% 			International Conference of}, pages 312--318. IEEE, 2014.

% 	\bibitem{hadash2018estimate}
% 	Guy Hadash, Einat Kermany, Boaz Carmeli, Ofer Lavi, George Kour, and Alon
% 	Jacovi.
% 	\newblock Estimate and replace: A novel approach to integrating deep neural
% 	networks with existing applications.
% 	\newblock {\em arXiv preprint arXiv:1804.09028}, 2018.

% \end{thebibliography}

\end{document}